\documentclass[prr,reprint,twocolumn,10pt,aps,longbibliography]{revtex4-2}

\usepackage{amsmath,amssymb,bm}
\usepackage{mathtools}
\usepackage{esvect}
\usepackage{booktabs}
\usepackage{multirow}
\usepackage{array}
\usepackage{upgreek}
\usepackage[svgnames]{xcolor}

\begin{document}

\title{Tracking a nonrelativistic charge with an array of Rydberg atoms}
\author{Akio Kawasaki}
\email{akio.kawasaki@aist.go.jp}
\affiliation{National Metrology Institute of Japan (NMIJ), National Institute of Advanced Industrial Science and Technology (AIST), 1-1-1 Umezono, Tsukuba, Ibaraki 305-8563, Japan}


\begin{abstract}
Charged particle tracking has played a key role in the development of particle physics, particularly for understanding phenomena involving short-lived particles precisely. As a platform for high-resolution charged particle tracking, an array of Rydberg atoms is theoretically analyzed. Utilizing the Ramsey sequence to accumulate the phase shift between the ground and a Rydberg excited state induced by the time-dependent Stark shift due to a moving charge, a nonrelativistic charged particle can be tracked with a precision of $\sim10$ nm, with a potential of higher resolution by optimizing reconstruction algorithm. Although a lot of technical difficulties need to be resolved, the proposed scheme can potentially serve as a charge tracker for relativistic charged particles as well. Also, this analysis can explain potential decoherence in the quantum computation with Rydberg atoms induced by residual ions and cosmic rays. 
\end{abstract}

\maketitle

\section{Introduction}
Visualizing trajectories of high-energy charged particles advanced the field of particle physics. Discoveries of short-lived elementary particles were made with different kinds of particle trackers, such as cloud chambers \cite{CTRWilson1911,CTRWilson1912}, bubble chambers \cite{PhysRev.87.665}, and wire chambers \cite{WireChamber}. One direction of the development of charged particle tracking is to scale up the tracking volume. Time projection chambers (TPCs) \cite{TPCReview} for dark matter searches have multi-ton scales \cite{PhysRevLett.129.161805,PhysRevLett.129.161804,LZFirstResult} and neutrino physics experiments now have 100-ton scale liquid argon TPCs \cite{ICARUSDetector,MicroBooNEDEtector,ProtoDUNEDetector} and 10-kiloton scale detectors are being constructed \cite{DUNEDetector}. Others aim at high position resolution. This is typically achieved by silicon detectors \cite{PDG}, and the state-of-the-art silicon strip \cite{AMSSiTracker,PAMELASiTracker,CMSSiTracker,BelleIISiTracker,ATLASSiTracker,SiTrackerReview} and pixel \cite{BelleIIDEPFET,ALICEMAPS,CMSPixel,ATLASPixel,LHCbPixel} detectors have position resolutions on the order of 1 $\upmu$m. Nuclear emulsions \cite{NIMA.554.247,NIMA.718.519,NIMA.556.80} have higher position resolutions. 
However, they are incapable of real-time data acquisition and are therefore sorted as a different kind of detectors. One of the limiting factors for the position resolution is the finite size of readout strips or pixels, which is currently on the order of 10 $\upmu$m. The position resolution $\sigma_x$ and the pitch $w$ of the strips or pixels has a relation 
\begin{equation}\label{EqSigmaW}
\sigma_x=w/\sqrt{12}
\end{equation} 
for digital readout \cite{SiTrackerReview}. If $w$ is reduced substantially, the position resolution can drastically increase. 

A system smaller than a semiconductor circuit with a good control is atoms. Pioneering experimental attempts to make use of atoms as a particle detector are performed with vapor cells \cite{DAMOPZeeman,DAMOPStark}. More complicated systems, such as traps for single atoms and ions, are ready for this application, thanks to advancements in quantum technologies in past decades. Particularly, atoms can be optically trapped without applying electric or magnetic fields. Recent developments in tweezer array systems enabled us to access single atoms independently \cite{Nature.551.579,PhysRevX.8.041055,PhysRevX.8.041054,PhysRevLett.122.143002}, and atoms can be configured in a defectless lattice of desired lattice arrangement and atom spacing \cite{Science.354.1021,OptExpress.27.2184,PhysRevA.102.063107,PhysRevApplied.19.034048}. Each atom in the array can be detected independently and thus functions as a single pixel for charge detection. The atom-atom distance is as small as a few micrometers and can go down to $O(100)$ nm with a quantum gas microscope setup \cite{Nature.462.74,Nature.467.68}. Naively, this can improve the position resolution of charge tracking by orders of magnitude. 

In this paper, a possibility of utilizing an array of Rydberg atoms as a charged particle tracker is theoretically discussed. As a starting point, tracking of a nonrelativistic charge is analyzed. A Rydberg state with a blockade radius slightly smaller than the distance between neighboring atoms enhances the sensitivity to the charge. The Ramsey sequence integrates the phase shift on the Rydberg state induced by the charge. With these configurations, atoms behave as substantially high-density pixels compared to a silicon tracker and the resolution of the tracking is improved by two orders of magnitude. The technical limitations of the proposed scheme and possible ways to extend it to a relativistic charge are also discussed. 

\begin{figure*}[!t]
    \includegraphics[width=2\columnwidth]{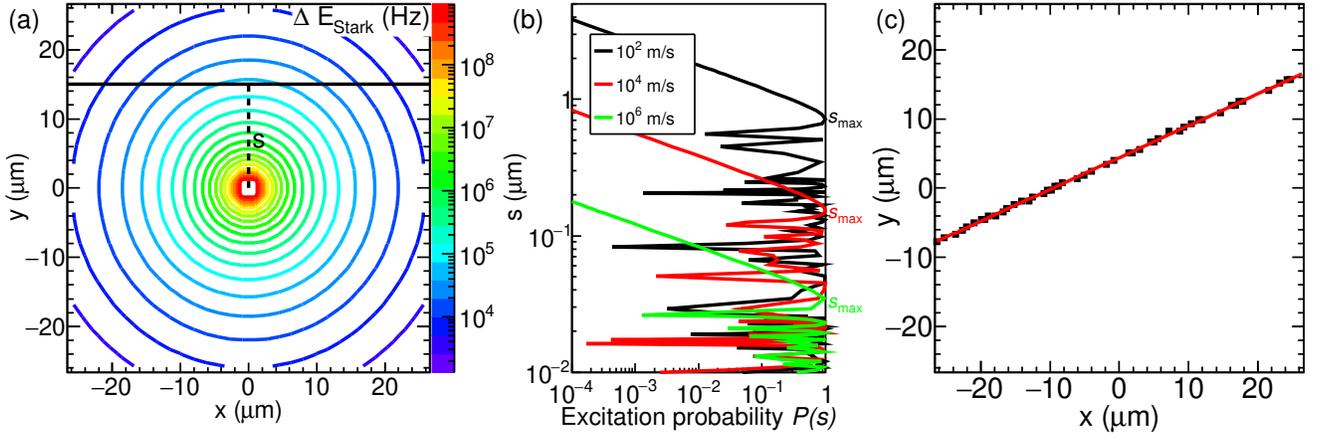}
    \caption{(a)Amount of the Stark shift $\Delta E_{ \rm Stark}$ for the $18P_{3/2}$ state of Cs induced by a $+e$ charge fixed at the origin. The black straight line shows a sample path for integrating the phase shift to obtain the transition rate. 
	(b) Probability $P(s)$ of an atom detected in the excited state for different impact parameters $s$. $s_{\rm max}$ is defined as the largest $s$ satisfying $P(s)=1$. Below $s_{\rm max}$, $P(s)$ undergoes fast oscillation between 0 and 1, and a finite step size for $s$ in the numerical calculation that generated the plot can be potentially larger than the period of the oscillation. 
	(c) A sample trajectory of a charge detected by the proposed Rydberg atom array. Black points are generated with $v=10^4$ m/s, $s=4$ $\upmu$m, and $\theta=2$ rad by a Monte Carlo simulation. The red line is the linear fit of the black points. Fitted values are $s=4.004(22)$ $\upmu$m and $\theta=2.0016(14)$. }
    \label{FigStark}
\end{figure*}

\section{Charge detection through Stark shift}\label{ExpSystem}
In the following discussion, atoms in an optical trap are assumed to form a two-dimensional square lattice with a lattice constant $d$. Two systems can realize this configuration. One is the quantum gas microscope setup, where atoms are confined in a single layer of a three-dimensional optical lattice. $d=\lambda/2$ is determined by the wavelength $\lambda$ of the trapping laser, and the detection system with a microscope objective has a single-atom resolution. Another setup is a tweezer array. This consists of hundreds of optical tweezers, each of which contains a single atom. Atoms can be positioned arbitrarily, where typical distance between atoms is a few micrometers, and both the detection and control can be performed independently for each atom. Atoms in a three-dimensional optical lattice allow three-dimensional tracking if they are imaged from two different directions. However, the highest reported resolution of 1.1 $\upmu$m \cite{PhysRevLett.120.103201} has not reached single-atom resolution yet, and thus we concentrate on the two-dimensional systems. 

A charged particle is detected by atoms through dc Stark shift $\Delta E_{\rm Stark}$ induced by the electric field $E$ generated by the charge. The lattice constant $d$ and state of atoms can be selected to maximize the sensitivity to the charge. Rydberg states with large principal quantum number $n$ are known as atomic states with large dc Stark shift, where the polarizability $\alpha$ scales $\alpha\sim n^7$. Large $n$ also enhances atom-atom interaction induced by the van der Waals interaction, which can potentially disturb the system. The strength of the van der Waals interaction is characterized by a constant $C_6$ that scales $C_6 \sim n^{11}/d^6$. To keep the amount of the van der Waals interaction constant, $n$ needs to satisfy $n\sim d^{6/11}$. A typical electric field generated by the charge at the atom closest to the charge scales $E\sim 1/d^2$. Therefore, overall typical sensitivity is depicted as $\Delta E_{\rm Stark}\sim \alpha E^2 \sim n^7/d^4 \sim d^{-2/11}$. This means smaller $d$ gives higher sensitivity to a charge. In practice, the smallest $d$ attainable in these systems is set by $\lambda$. In the following discussion, a quantum gas microscope system with $\lambda=1064$ nm light is assumed.

To be more specific, a $100\times100$ square lattice of Cs atoms in a $P_{3/2}$ state with $d=532$ nm is considered. Atoms are located at the lattice points $(x,y)=(d(2i+1)/2, d(2j+1)/2)$, where $i,j=-50,49,...,49$ are integers. The $84P_{3/2}$ state for Cs has a blockade radius of $r_{\rm B}=8.99$ $\upmu$m and polarizability of 2.38 MHz/(V/m)$^2$ \cite{PhysRevA.89.033416}. With quantum defect $\delta=3.559$ \cite{PhysRevA.26.2733} and an assumption of a Rabi frequency being 10 kHz, the maximum $n$ for the $P_{3/2}$ state with $r_{\rm B}<d$ is $n=18$, which gives $\alpha=14.3$ Hz/(V/m)$^2$ and $r_{\rm B}=386$ nm. A  charge of $+e$ fixed at the origin induces  
\begin{equation}
\Delta E_{\rm Stark}(x,y)=-\alpha E(x,y)^2=-\frac{1}{16\pi^2\varepsilon_0^2}\frac{\alpha e^2}{(x^2+y^2)^2}.
\end{equation}
Figure \ref{FigStark}(a) shows that $\Delta E_{\rm Stark}>100$ MHz at small distances, which is well detectable. 

When the charge moves at a velocity $v$, $\Delta E_{\rm Stark}$ changes over time. Responses other than $\Delta E_{\rm Stark}$, such as transitions to different Rydberg states and ionization, can also happen. $\Delta E_{\rm Stark}$ is first analyzed as the least destructive and therefore the most sensitive response. Other responses are discussed in Section \ref{SecIII}. The moving charge is assumed to be under uniform linear motion, with an impact parameter $s$ with respect to an atom. Without losing generality, the relative position of the charge from an atom can be assumed to move on the black line in Fig. \ref{FigStark} (a), and hence $(x(t), y(t))=(vt,s)$. The infinitesimal phase shift induced on the Rydberg state by the Stark shift can be integrated over time, resulting in the overall phase shift $\Delta \phi$ between the Rydberg state and the ground state. 
\begin{equation}
\Delta \phi(s)= \int^{\infty}_{-\infty} dt(\omega-\omega_0(t))t= \int^{\infty}_{-\infty} dt \frac{\Delta E_{\rm Stark}(vt,s)t}{\hbar},
\end{equation}
where $\omega$ and $\omega_0$ are frequencies for a local oscillator and the atomic resonant frequency from the ground state $|g \rangle$ to the Rydberg excited state $|e \rangle$, respectively.

To detect $\Delta \phi$, the Ramsey sequence, where the $1:1$ superposition state between $|g \rangle$ and $|e \rangle$ records the accumulated phase shift between $|g \rangle$ and $|e \rangle$ over an interrogation time $\tau$, is suitable. $\omega$ is assumed to be stable over $\tau$ and tuned at unperturbed atomic resonance, whereas $\Delta E_{\rm Stark}$ modulates $\omega_0$. The interrogation time needs to start prior to the arrival of the charge and to end after the charge passed by to ensure that a charge affects the whole atom array system in the final state. 

The minimum $\tau$ required for the sequence is determined by the slowest charge. A charge with the thermal velocity $v\sim100$ m/s of a residual gas ion needs 530 ns to pass the array. This is shorter than the lifetime of the Rydberg state ($4.0$ $\upmu$s for the $15P_{3/2}$ state in Cs \cite{PhysRevA.79.052504} and therefore slightly longer for the $18P_{3/2}$ state). One of the slowest charges conceivable in the context of nuclear and particle physics is an $\alpha$ particle with 1-MeV energy, whose velocity $3.1 \times 10^6$~m/s is substantially faster than the thermal ion. Thus the interrogation time can be arranged to start much earlier than the arrival of a charge and to end much later than when the charge leaves the system. Note that the start of the interrogation time can be triggered by an arrival of a bunch of particles, if the experiment is based on the particle beam coming from an accelerator. The trigger rate is assumed to be $\lesssim 1$ Hz, because it typically takes $\lesssim 1$ s to prepare atoms. 

The phase shift imprinted in the atomic state is detected by converting the phase shift to the population difference by a $\pi/2$ pulse. The final $\pi/2$ pulse has a phase offset of $\pi$ compared to the initial $\pi/2$ pulse to turn atoms not affected by the charge back to $|g \rangle$. This suppresses any potential excitation of atoms without interaction between the charge. Atoms affected by the charge have the excitation probability $P(s)=\sin^2 (\Delta\phi(s)/2)$. Once the state is projected to $|g \rangle$ or $|e \rangle$, the detection efficiency can be assumed to be 100\%; for example, the atoms in $|e \rangle$ can be ionized and atoms remaining in $|g \rangle$ can be detected. The detection of the atoms in $|g \rangle$ is typically performed by shining imaging light and cooling light simultaneously \cite{PhysRevLett.122.143002} or alternatingly \cite{PhysRevLett.122.173201} for a few tens of milliseconds, where atoms can be detected with probability more than 99.99\%, with their lifetime in the trap being at most 10 s. 

$P(s)$ is shown in Fig. \ref{FigStark}(b). $s_{\rm max}$ is defined as the largest $s$ satisfying $P(s)=1$. $P(s)$ decreases monotonically above $s_{\rm max}$, and it rapidly oscillates between 0 and 1 below $s_{\rm max}$, resulting in the average excitation probability of 0.5. The region with $s<s_{\rm max}$ can leave atoms in $|e \rangle$, which are recorded as hits. 

\section{Charge detection through ionization and transitions}\label{SecIII}

\begin{figure}[!t]
    \includegraphics[width=1\columnwidth]{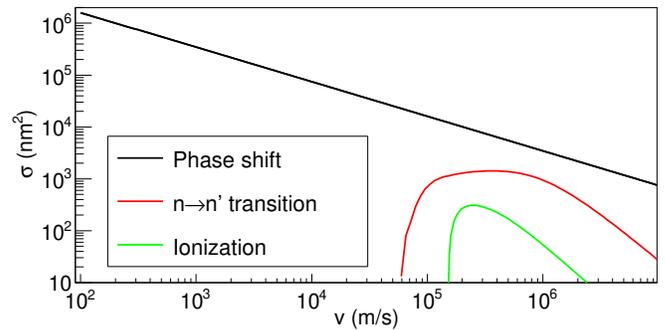}
    \caption{Cross sections $\sigma$ for the $18P_{3/2}$ state in Cs atoms by impacts of electrons with velocity $v$. Phase shift corresponds to the detection scheme through Stark shift described in the main text. $n \rightarrow n'$ transition indicates the sum of the transitions of all Rydberg states with ${n^*}'-n^*>1$. Ionization shows the cross section for the ionization of the Rydberg electron. }
    \label{FigSigmas}
\end{figure}

Ionization and transitions to different Rydberg states can also be induced by a moving charge. Such interactions are previously studied both experimentally and theoretically \cite{PhysRep.250.95,RydStateAtMol,JPhysB.42.022001}. Figure \ref{FigSigmas} compares the cross sections of three processes for different $v$. Cross sections are estimated from empirically obtained equations for ionization and transition to different states \cite{PhysRevA.22.940} (see Appendix for the equations deriving the plot) and conservatively by $\pi s_{\rm max}^2$ for the Stark shift. For both ionization and transitions to different Rydberg states, a quantum defect of $\delta=3.559$ for the $18P_{3/2}$ state is included to calculate the energy level of the initial Rydberg state. To represent the final Rydberg states, $\delta=0$ is assumed, because the final state can have large azimuthal quantum number $l$ \cite{JPhysB.12.427} where $\delta=0$.

For the states satisfying ${n^*}'-n^*<1$, where $n^*$ (${n^*}'$) is the principal quantum number compensated by $\delta$ for the initial (final) state, the empirical equations break down and thus the $n\rightarrow n'$ transition in Fig. \ref{FigSigmas} excludes such transitions. Cross sections for the transitions to different $l$ within the same $n$ states are experimentally measured for Na with $v\sim 10^5$~m/s \cite{PhysRevA.24.1286} and theoretically analyzed \cite{PhysRep.250.95}, which are not reproduced by the empirical equations. Extrapolation of Ref. \cite{PhysRevA.24.1286} data to the $n^*$ for the $18P_{3/2}$ state in Cs with the $n$ scaling of $N\sim \sigma^{5.12}$ gives $1.8\times 10^5$ nm$^2$. Based on the note in Ref. \cite{PhysRevA.24.1286} that at least an order of magnitude smaller cross section is expected for an initial state with greater isolation in the energy diagram, the cross section for ${n^*}'-n^*<1$ transitions is expected to be at most $10^4$ nm$^2$, which is similar to the cross section for the phase shift detection scheme. 
 
Excitation or ionization of Rydberg atoms by half-cycle pulses \cite{PhysRevA.51.3370,JPhysB.42.022001} is equivalent to those of Rydberg atoms by a moving charge, because the moving charge also induces a unipolar short-pulse electric field. The required electric field for ionizing 50 \% of an $n=18$ state for Cs by a half-cycle pulse is estimated to be 50 kV/cm by extrapolating an experimental result in Ref. \cite{PhysRevA.51.3370}. Such a peak electric field can be generated when $s<17$ nm, resulting in a cross section of $\sigma_{\rm HCP}=\pi s_{\rm max}^2=9\times 10^2$ nm$^2$. The pulse length requirement that the pulse needs to be shorter than a Keplerian period of the Rydberg electron requires $v\geq 7.4\times 10^4$~m/s. These numbers are on the order of magnitude same as the ionization case in Fig. \ref{FigSigmas}, and thus justify the validity of the estimates in Fig. \ref{FigSigmas}.

Ionization and transitions to different energy levels happen only when the energy of the incident electron is above the transition energy. Above the maximum, these cross sections decay with the scaling $\sigma\sim v^{-4}$. The plot shows both the ionization and transition to different $n$ states has smaller cross sections than that for the phase-shift measurement. These analyses agree with the intuition that the phase shift is the least inelastic and thus the most sensitive to the existence of a moving charge. For all interactions, as far as a hit is defined as an atom not remaining in $|g \rangle$, the basic detection method still can be the same as the detection by the phase shift.

\section{Result of the simulation}
To see the performance of the tracking, Monte Carlo simulations are performed. A charge of $+e$ is assumed to fly on a random line at a fixed velocity $v$ ranging from 10 m/s to $10^7$ m/s. The largest $v$ is chosen to keep all calculations nonrelativistic. The random line is first generated by selecting a distance $l$ of the line from the origin and its angle $\theta$ between the $x$ axis from the uniform distribution. Next, the final states for atoms are calculated according to the procedure described in Sec. \ref{ExpSystem}. Atoms in $|e \rangle$ are recorded as hits, and the hit data are used to reconstruct the trajectory. 

\begin{figure}[!t]
    \includegraphics[width=1\columnwidth]{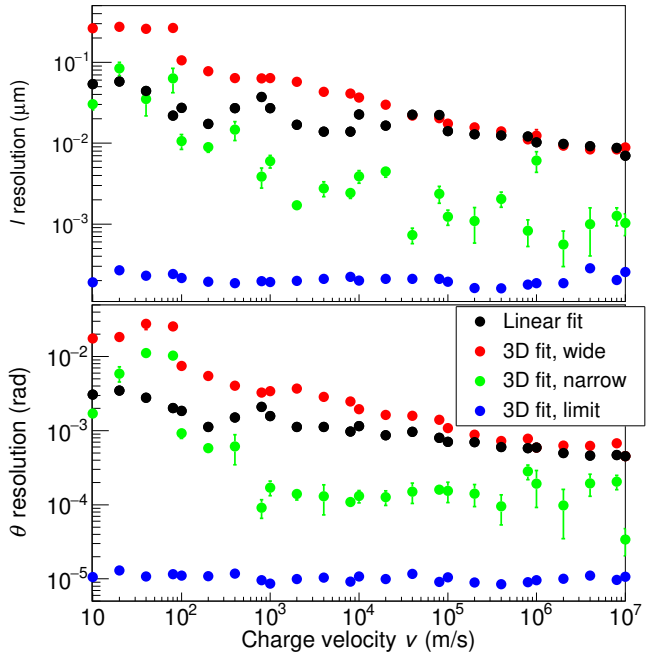}
    \caption{Resolution for the charged particle tracking: fits are performed for single linear trajectories. Parameters are the distance $l$ from the origin and the angle of the trajectory $\theta$ against the $x$ axis, whose resolutions are plotted in the top and bottom half, respectively. Black points show the resolution for the linear fit. Other points are related to the 3D fit. Green and red points are the standard deviations for narrow and wide Gaussian derived by a fit with two Gaussians. Blue points are the ultimate resolution for the 3D fit.}
    \label{FigVvsWidthRT}
\end{figure}

The simplest way of reconstruction under the assumptions is a linear fit of the hits. To assure the uniform evaluation of the fit, the fitted line is also parametrized by $l$ and $\theta$: $y=-x/\tan\theta+l/\sin\theta$. An example shown in Fig. \ref{FigStark}(c) demonstrates a good linear fit for the atoms in $|e \rangle$. The simulation is run over 1000 trials for each $v$, and the standard deviation for the distribution of the difference of the fitted value from the true value is regarded as the resolution. The resolution, shown as black points in Fig. \ref{FigVvsWidthRT}, is on the order of 10 nm for $l$ and $10^{-3}$ for $\theta$. The angular resolution also results in at most an $O(10)$ nm position uncertainty, as the overall size of the system is 53.2 $\mu$m. The resolution is slightly higher for large $v$, presumably because $s_{\rm max}$ is smaller. 

\begin{figure}[!t]
    \includegraphics[width=1\columnwidth]{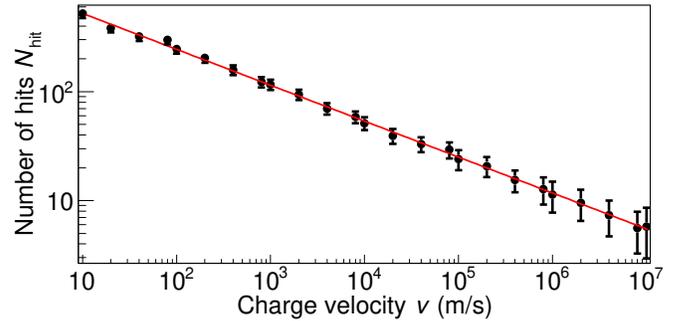}
    \caption{Number of total hits $N_{\rm hit}$ in the array for different charge velocities $v$. The red line is a fit by $N_{\rm hit}=p_0 v^{p_1}$, with the fitted values of $p_0=1117(65)$ and $p_1=-0.3297(82)$.}
    \label{FigVvsNhit}
\end{figure}

Dependence of $s_{\rm max}$ on $v$ can be used for a velocity estimate. Figure \ref{FigVvsNhit} shows that the total number of hits $N_{\rm hit}$ scales to $v^{-1/3}$. Even with the smallest uncertainty in $N_{\rm hit}$ of 8.6\% at $v=10$ m/s, the relative energy resolution is $\sim50$\%, which is at most to the extent of an order-of-magnitude estimate. Precise determination needs to be performed with different types of detectors, such as a calorimeter, located downstream. 

If $v$ is determined by a different detector with high enough precision, resolution of the reconstruction can be improved by more sophisticated fit than the linear fit. To test this, 0 and 1 are first assigned to atoms in $|g \rangle$ and $|e \rangle$, respectively, to obtain a three-dimensional plot (and thus this fit is called 3D fit). This three-dimensional plot is regarded as the data $P_{ij}$. The fitting function is the line $y=-x/\tan\theta+l/\sin\theta$. 
To perform the fit, the expected excitation probaility $P_{ij}(l,\theta)$ is calculated for each atom based on P(s), where s is the distance of the atom labeled by (i,j) from the line. The uncertainty $\sigma_{ij}$ for $P_{ij}(l,\theta)$ is calculated as that for binomial distribution. The best fit is obtained as the combination of $l$ and $\theta$ minimizing the following $\chi^2$.
\begin{equation}
\chi^2=\sum_{i,j}\frac{(P_{ij}-P_{ij}(l,\theta))^2}{\sigma_{ij}^2} \nonumber
\end{equation}

Reconstruction by the 3D fit is more precise than that by the linear fit only when the initial values are chosen properly; the distribution for the difference of the fitted value from the true value is fitted well with two Gaussians with different standard deviations besides a small number of long tails where the reconstruction did not converge correctly. The resolution for the broad Gaussian is at best comparable to the linear fit for large $v$, as shown in Fig. \ref{FigVvsWidthRT}. However, for the narrow Gaussian, the uncertainty is at most an order of magnitude smaller than the linear fit. Ultimate resolution when the initial values are set to the true values is higher than that of the linear fit by two orders of magnitude. The sophisticated fitting algorithm as well as $v$ estimate can enhance the resolution of the charged particle tracking further compared to the algorithm shown here.

\section{Discussion}

Although the resolution around 10 nm by the linear fit is two orders of magnitude higher than that for the state-of-the-art silicon trackers, the atom array system has various technical limitations. A crucial problem for the application to high-energy physics is the limited sensitivity for large $v$. $N_{\rm hit}$ is only a few at $v=10^7$ m/s and decreases to zero for larger $v$. The larger lattice constant $d$ can enhance the maximum trackable velocity proportional to $d$ when a Rydberg state with larger $n$ is selected to increase the polarizability. Experimentally, $d=9$ $\upmu$m is achieved \cite{PhysRevX.8.041055}, and therefore covering $v=3\times 10^8$ m/s charge is possible. A disadvantage for large $d$ is the reduced resolution $\sigma \sim d$ due to Eq. (\ref{EqSigmaW}). Also, the total number of atoms in an array potentially needs to be reduced, due to the finite field of view for the imaging system. It should be noted that the electric field needs to be calculated in a relativistic manner for a relativistic charge, with the magnetic field coming into consideration as well. 

The interaction between the charge and atoms can also be enhanced by a larger amount of charge. Such enhancement works for heavy nuclei, whose charge can exceed $+100e$. Also, beam position monitoring for a bunched particle beam can be a good application.

The second major problem is the limited trackability due to the two-dimensional structure. When a charged particle flies in a three-dimensional direction, only a part of the trajectory where the charge is close enough to atoms can be tracked. To partially allow three-dimensional tracking, a second tweezer array can be located perpendicular to the first one. A three-dimensional optical lattice can potentially have a single-atom resolution if a microscope objective for imaging is translated axially to image multiple layers. 

Optical components for the atom array need to be carefully arranged. On one hand, for an atom array system, it is essential for optics components for trap formation and imaging, such as a microscope objective, mirrors, and a camera, to be located close to the array. On the other hand, as a particle detector, it is desired to have the smallest amount of material on the path of particles. Particularly, for collider experiments, a charge tracker with fine pixels is located at the inner-most layer of the detectors, outside of which as small as possible amount of matter is desired before charged particles reach outer-layer detectors. The optics components have to be located to avoid the path for the particles as much as possible. Note that the optical system for an initial atom trap does not have to be close to the atom array system; they can be conveyed on a moving optical lattice from somewhere farther. 

Some electromagnetic disturbance to the atomic system needs to be carefully managed. The static electric field has to be weaker than the threshold for the field ionization. The threshold is on the order of 1 kV/cm for the $18P_{3/2}$ state in Cs, estimated from the case of Na \cite{PhysRevA.16.1098}. To keep high-voltage sources away from the region for the Rydberg atoms, other detectors need to be properly arranged. For example, silicon trackers do not have as high voltage as TPCs. When a TPC or other type of drift chamber that can have a $\sim100$ kV electrode is located next to the Rydberg atom system, the ground plane should face the atoms. With the dc electric field kept below the ionization threshold, constant backgrounds can be calibrated in advance and compensated accordingly. Transient electromagnetic backgrounds generated by accelerator components close to the detectors can be a major problem. Also, if the atom array is close to the path for the colliding particle beams, the electromagnetic fields from the particle beam can disturb the atomic energy levels. To avoid them, the Rydberg atom system needs to be located far from the main particle beam. Because Rydberg atoms can be formed in strong magnetic fields \cite{PhysRevA.87.012505,PhysRep.484.181}, a constant or predictable amount of magnetic field does not disturb the system. Proper calibration and compensation by the laser frequency to manipulate the atomic state can manage the effect due to the magnetic field. 

Based on these concerns regarding practical implementation of the proposed system, the first  implementation of an atom array as a charged particle tracker would be to an experiment of radioactive heavy nuclei synthesis. Such experiments often have a beam separator to extract desired nuclei. Separating the nuclei that are not of one's interest reduces the event rate, suppressing the disturbance due to the particle beams that did not have any reactions, including the photons emitted by the main particle beam. Because particles coming out of the separator are focused and oriented to a specific direction, partial three-dimensional tracking is possible by putting two atom arrays intersecting on the average trajectory of the incident particles, without putting any extra matters on the path of outgoing particles. More specifically, define the approximate path for the charged particles as the $+z$ axis. Two atom arrays should be formed on the $x-z$ plane and the $y-z$ plane. Because the atom array on the $x-z$ ($y-z$) plane requires optics components for forming the array and detection along the $y$ ($x$) axis, the surrounding optics components do not conflict with each other, the charged particles, or other detectors at downstream in the $+z$ direction. The nuclei to be detected can have a large charge of $\sim+100e$, which enhances the sensitivity of the atom array. The precise tracking benefits the lifetime measurement by time of flight. 

Many-body effects other than a Rydberg blockade for two Rydberg atoms also need to be considered. The overall shift by van der Waals interaction with other atoms in the two-dimensional array is numerically calculated to be 4.66 times larger than the two-atom case. This increases $r_B$ by a factor of 1.29, but $r_B<d$ is still satisfied for the setup discussed in Sec. \ref{ExpSystem}. The interaction is also affected by the edge effect. The amount is at most on the order of $0.1\Gamma$ \cite{PhysRevLett.127.013401}, where $\Gamma\sim100$ kHz \cite{PhysRevA.79.052504} is the linewidth of the transition, and thus the shift is expected to be at most comparable to the assumed Rabi frequency of 10 kHz. Even if the shift is larger than 10 kHz, all atoms in an optical lattice can be excited to a Rydberg state by a pulsed laser, and a Ramsey sequence for such atoms is experimentally demonstrated \cite{NatPhoton.16.724,2201.09590}.

The analysis here can also be useful to estimate the potential decoherence of a quantum computer based on Rydberg atoms. If a residual gas or a cosmic ray passing by a qubit is charged, the energy level of the qubit is perturbed by these particles. An ion with a thermal velocity in the room temperature, at most $10^3$ m/s, is well detectable by a Rydberg atom. Cosmic rays can also affect Rydberg atoms, if a cosmic ray is nonrelativistic or if $n$ for the Rydberg state is large. The phase shift induced by these charged particles between the ground state and the Rydberg state, or another qubit, can induce an error in calculations. 

\section{Conclusion}
To summarize, a possibility of making use of an array of Rydberg atoms for tracking a nonrelativistic charge is investigated. By integrating the phase shift induced by the Stark shift with a Ramsey sequence, a moving charge can be tracked with a resolution of $\sim10$ nm, which is two orders of magnitude higher than that for the state-of-the-art silicon tracker. The resolution can be potentially improved further with an optimized fitting procedure. The analysis can be used for estimating an impact of a charge flying by a qubit made of a Rydberg atom.

\begin{acknowledgments}
This work is supported by JSPS KAKENHI Grant No. JP22H01161 and JST FOREST Grant No. JPMJFR212S. The author thanks Alexander Fieguth, Takeshi Fukuhara, and Takafumi Tomita for insightful discussion. 
\end{acknowledgments}

\appendix
\section{Empirical electron-impact cross sections}
The equation to estimate the cross sections for ionization is 
\begin{equation}
\sigma^I_n(E)=\frac{\pi \alpha_0^2 \hbar^2 c^2}{E+3.25I_n} \left( \frac{5}{3I_n} - \frac{1}{E} -\frac{2}{3} \frac{I_n}{E^2} \right),
\end{equation}
where $E$ is the energy of the incident electron, $I_n=R_{\infty}/n^2$ is the ionization energy from the initial Rydberg state with principal quantum number $n$, $\alpha_0$ is the fine-structure constant, $\hbar$ is the reduced Planck constant, and $c$ is the speed of light \cite{PhysRevA.22.940}.

The transition from $n$ state to $n'$ state has the following cross section \cite{PhysRevA.22.940}:
\begin{equation}
\sigma_{nn'}(E)=\frac{\pi a_0^2 R_{\infty}}{E+\alpha_{nn'}}\left( A_{nn'} \ln \left( \frac{E}{R_{\infty}} +\beta_{nn'} \right) +B_{nn'} \right),
\end{equation}
where $R_{\infty}$ is the Rydberg constant and $a_0$ is the Bohr radius. $A_{nn'}$, $B_{nn'}$, $\alpha_{nn'}$, $\beta_{nn'}$ are described in the following way: 
\begin{eqnarray}
A_{nn'}&=& \frac{2R_{\infty}}{E} f_{nn'} \nonumber \\
B_{nn'}&=& \frac{1}{n'^3} \left[ \frac{2R_{\infty}}{E_{nn'}} \right)^2 \left( 1+\frac{4}{3} \frac{I_n}{E_{nn'}}+ \frac{b_n}{n} \left(\frac{I_n}{E_{nn'}}\right)^2 \right] \nonumber \\
\alpha_{nn'} &=& \frac{8+23s^2/n^2}{8+1.1n\Delta n+0.8/\Delta n^2+0.4(n^3/\Delta n)^{1/2}|\Delta n-1|} \nonumber \\
\beta_{nn'}  &=& \exp\left(-\frac{B_{nn'}}{A_{nn'}}\right)-0.4\frac{E_{nn'}}{R_{\infty}}\nonumber 
\end{eqnarray}
with $f_{nn'}$ being the oscillator strength between the $n$ and $n'$ state, $\Delta n= n'-n$, $E_{nn'}=R_{\infty}(1/n^2-1/n'^2) $ being the energy difference between the $n$ and $n'$ state, and 
\begin{equation}
b_n=1.41 \ln n - 0.7 - \frac{0.51}{n} + \frac{1.16}{n^2} -\frac{0.55}{n^3}. \nonumber 
\end{equation}
To estimate $f_{nn'}$, the following equations are used \cite{AstrophysJ.174.227}:
\begin{eqnarray}
f_{nn'}&=&\frac{32}{3\sqrt{2} \pi} \frac{n}{n'^3} x^{-3} g(n,x), \nonumber \\
x&=&\frac{E_{nn'}}{I_n}=1-\left(\frac{n}{n'}\right)^2, \nonumber \\
g(n,x)&=&g_0(n) + g_1(n) \frac{1}{x} + g_2(n) \frac{1}{x^2}, \nonumber \\
g_0(n) &=& 0.9935+\frac{0.2328}{n}-\frac{0.1296}{n^2} \nonumber \\
g_1(n) &=& -\frac{1}{n}\left( 0.6282-\frac{0.5598}{n}+\frac{0.5299}{n^2} \right) \nonumber \\
g_2(n) &=& \frac{1}{n^2}\left( 0.3887-\frac{1.181}{n}+\frac{1.470}{n^2} \right) \nonumber 
\end{eqnarray}

\bibliography{RydbergCTRev}

\begin{thebibliography}{58}%
\makeatletter
\providecommand \@ifxundefined [1]{%
 \@ifx{#1\undefined}
}%
\providecommand \@ifnum [1]{%
 \ifnum #1\expandafter \@firstoftwo
 \else \expandafter \@secondoftwo
 \fi
}%
\providecommand \@ifx [1]{%
 \ifx #1\expandafter \@firstoftwo
 \else \expandafter \@secondoftwo
 \fi
}%
\providecommand \natexlab [1]{#1}%
\providecommand \enquote  [1]{``#1''}%
\providecommand \bibnamefont  [1]{#1}%
\providecommand \bibfnamefont [1]{#1}%
\providecommand \citenamefont [1]{#1}%
\providecommand \href@noop [0]{\@secondoftwo}%
\providecommand \href [0]{\begingroup \@sanitize@url \@href}%
\providecommand \@href[1]{\@@startlink{#1}\@@href}%
\providecommand \@@href[1]{\endgroup#1\@@endlink}%
\providecommand \@sanitize@url [0]{\catcode `\\12\catcode `\$12\catcode
  `\&12\catcode `\#12\catcode `\^12\catcode `\_12\catcode `\%12\relax}%
\providecommand \@@startlink[1]{}%
\providecommand \@@endlink[0]{}%
\providecommand \url  [0]{\begingroup\@sanitize@url \@url }%
\providecommand \@url [1]{\endgroup\@href {#1}{\urlprefix }}%
\providecommand \urlprefix  [0]{URL }%
\providecommand \Eprint [0]{\href }%
\providecommand \doibase [0]{https://doi.org/}%
\providecommand \selectlanguage [0]{\@gobble}%
\providecommand \bibinfo  [0]{\@secondoftwo}%
\providecommand \bibfield  [0]{\@secondoftwo}%
\providecommand \translation [1]{[#1]}%
\providecommand \BibitemOpen [0]{}%
\providecommand \bibitemStop [0]{}%
\providecommand \bibitemNoStop [0]{.\EOS\space}%
\providecommand \EOS [0]{\spacefactor3000\relax}%
\providecommand \BibitemShut  [1]{\csname bibitem#1\endcsname}%
\let\auto@bib@innerbib\@empty
\bibitem [{\citenamefont {Wilson}(1911)}]{CTRWilson1911}%
  \BibitemOpen
  \bibfield  {author} {\bibinfo {author} {\bibfnamefont {C.~T.~R.}\
  \bibnamefont {Wilson}},\ }\bibfield  {title} {\bibinfo {title} {On a method
  of making visible the paths of ionising particles through a gas},\ }\href
  {https://doi.org/10.1098/rspa.1911.0041} {\bibfield  {journal} {\bibinfo
  {journal} {Proc. R. Soc. Lond. A}\ }\textbf {\bibinfo {volume} {85}},\
  \bibinfo {pages} {285} (\bibinfo {year} {1911})}\BibitemShut {NoStop}%
\bibitem [{\citenamefont {Wilson}(1912)}]{CTRWilson1912}%
  \BibitemOpen
  \bibfield  {author} {\bibinfo {author} {\bibfnamefont {C.~T.~R.}\
  \bibnamefont {Wilson}},\ }\bibfield  {title} {\bibinfo {title} {On an
  expansion apparatus for making visible the tracks of ionising particles in
  gases and some results obtained by its use},\ }\href
  {https://doi.org/10.1098/rspa.1912.0081} {\bibfield  {journal} {\bibinfo
  {journal} {Proc. R. Soc. Lond. A}\ }\textbf {\bibinfo {volume} {87}},\
  \bibinfo {pages} {277} (\bibinfo {year} {1912})}\BibitemShut {NoStop}%
\bibitem [{\citenamefont {Glaser}(1952)}]{PhysRev.87.665}%
  \BibitemOpen
  \bibfield  {author} {\bibinfo {author} {\bibfnamefont {D.~A.}\ \bibnamefont
  {Glaser}},\ }\bibfield  {title} {\bibinfo {title} {Some effects of ionizing
  radiation on the formation of bubbles in liquids},\ }\href
  {https://doi.org/10.1103/PhysRev.87.665} {\bibfield  {journal} {\bibinfo
  {journal} {Phys. Rev.}\ }\textbf {\bibinfo {volume} {87}},\ \bibinfo {pages}
  {665} (\bibinfo {year} {1952})}\BibitemShut {NoStop}%
\bibitem [{\citenamefont {Charpak}\ \emph {et~al.}(1968)\citenamefont
  {Charpak}, \citenamefont {Bouclier}, \citenamefont {Bressani}, \citenamefont
  {Favier},\ and\ \citenamefont {\v{C}. Zupan\v{c}i\v{c}}}]{WireChamber}%
  \BibitemOpen
  \bibfield  {author} {\bibinfo {author} {\bibfnamefont {G.}~\bibnamefont
  {Charpak}}, \bibinfo {author} {\bibfnamefont {R.}~\bibnamefont {Bouclier}},
  \bibinfo {author} {\bibfnamefont {T.}~\bibnamefont {Bressani}}, \bibinfo
  {author} {\bibfnamefont {J.}~\bibnamefont {Favier}},\ and\ \bibinfo {author}
  {\bibnamefont {\v{C}. Zupan\v{c}i\v{c}}},\ }\bibfield  {title} {\bibinfo
  {title} {The use of multiwire proportional counters to select and localize
  charged particles},\ }\href
  {https://doi.org/https://doi.org/10.1016/0029-554X(68)90371-6} {\bibfield
  {journal} {\bibinfo  {journal} {Nucl. Instrum. Methods}\ }\textbf {\bibinfo
  {volume} {62}},\ \bibinfo {pages} {262} (\bibinfo {year} {1968})}\BibitemShut
  {NoStop}%
\bibitem [{\citenamefont {Hilke}(2010)}]{TPCReview}%
  \BibitemOpen
  \bibfield  {author} {\bibinfo {author} {\bibfnamefont {H.~J.}\ \bibnamefont
  {Hilke}},\ }\bibfield  {title} {\bibinfo {title} {Time projection chambers},\
  }\href {https://doi.org/10.1088/0034-4885/73/11/116201} {\bibfield  {journal}
  {\bibinfo  {journal} {Rep. Prog. Phys.}\ }\textbf {\bibinfo {volume} {73}},\
  \bibinfo {pages} {116201} (\bibinfo {year} {2010})}\BibitemShut {NoStop}%
\bibitem [{\citenamefont {Aprile}\ \emph {et~al.}(2022)\citenamefont {Aprile},
  \citenamefont {Abe}, \citenamefont {Agostini}, \citenamefont
  {Ahmed~Maouloud}, \citenamefont {Althueser}, \citenamefont {Andrieu},
  \citenamefont {Angelino}, \citenamefont {Angevaare}, \citenamefont {Antochi},
  \citenamefont {Antón~Martin} \emph {et~al.}}]{PhysRevLett.129.161805}%
  \BibitemOpen
  \bibfield  {author} {\bibinfo {author} {\bibfnamefont {E.}~\bibnamefont
  {Aprile}}, \bibinfo {author} {\bibfnamefont {K.}~\bibnamefont {Abe}},
  \bibinfo {author} {\bibfnamefont {F.}~\bibnamefont {Agostini}}, \bibinfo
  {author} {\bibfnamefont {S.}~\bibnamefont {Ahmed~Maouloud}}, \bibinfo
  {author} {\bibfnamefont {L.}~\bibnamefont {Althueser}}, \bibinfo {author}
  {\bibfnamefont {B.}~\bibnamefont {Andrieu}}, \bibinfo {author} {\bibfnamefont
  {E.}~\bibnamefont {Angelino}}, \bibinfo {author} {\bibfnamefont {J.~R.}\
  \bibnamefont {Angevaare}}, \bibinfo {author} {\bibfnamefont {V.~C.}\
  \bibnamefont {Antochi}}, \bibinfo {author} {\bibfnamefont {D.}~\bibnamefont
  {Antón~Martin}}, \emph {et~al.} (\bibinfo {collaboration} {XENON
  Collaboration}),\ }\bibfield  {title} {\bibinfo {title} {Search for new
  physics in electronic recoil data from {XENONnT}},\ }\href
  {https://doi.org/10.1103/PhysRevLett.129.161805} {\bibfield  {journal}
  {\bibinfo  {journal} {Phys. Rev. Lett.}\ }\textbf {\bibinfo {volume} {129}},\
  \bibinfo {pages} {161805} (\bibinfo {year} {2022})}\BibitemShut {NoStop}%
\bibitem [{\citenamefont {Zhang}\ \emph {et~al.}(2022)\citenamefont {Zhang},
  \citenamefont {Abdukerim}, \citenamefont {Bo}, \citenamefont {Chen},
  \citenamefont {Chen}, \citenamefont {Chen}, \citenamefont {Cheng},
  \citenamefont {Cheng}, \citenamefont {Cui}, \citenamefont {Fan} \emph
  {et~al.}}]{PhysRevLett.129.161804}%
  \BibitemOpen
  \bibfield  {author} {\bibinfo {author} {\bibfnamefont {D.}~\bibnamefont
  {Zhang}}, \bibinfo {author} {\bibfnamefont {A.}~\bibnamefont {Abdukerim}},
  \bibinfo {author} {\bibfnamefont {Z.}~\bibnamefont {Bo}}, \bibinfo {author}
  {\bibfnamefont {W.}~\bibnamefont {Chen}}, \bibinfo {author} {\bibfnamefont
  {X.}~\bibnamefont {Chen}}, \bibinfo {author} {\bibfnamefont {Y.}~\bibnamefont
  {Chen}}, \bibinfo {author} {\bibfnamefont {C.}~\bibnamefont {Cheng}},
  \bibinfo {author} {\bibfnamefont {Z.}~\bibnamefont {Cheng}}, \bibinfo
  {author} {\bibfnamefont {X.}~\bibnamefont {Cui}}, \bibinfo {author}
  {\bibfnamefont {Y.}~\bibnamefont {Fan}}, \emph {et~al.} (\bibinfo
  {collaboration} {PandaX Collaboration}),\ }\bibfield  {title} {\bibinfo
  {title} {Search for light fermionic dark matter absorption on electrons in
  {PandaX-4T}},\ }\href {https://doi.org/10.1103/PhysRevLett.129.161804}
  {\bibfield  {journal} {\bibinfo  {journal} {Phys. Rev. Lett.}\ }\textbf
  {\bibinfo {volume} {129}},\ \bibinfo {pages} {161804} (\bibinfo {year}
  {2022})}\BibitemShut {NoStop}%
\bibitem [{\citenamefont {Aalbers}\ \emph {et~al.}(2023)\citenamefont
  {Aalbers}, \citenamefont {Akerib}, \citenamefont {Akerlof}, \citenamefont
  {Al~Musalhi}, \citenamefont {Alder}, \citenamefont {Alqahtani}, \citenamefont
  {Alsum}, \citenamefont {Amarasinghe}, \citenamefont {Ames}, \citenamefont
  {Anderson} \emph {et~al.}}]{LZFirstResult}%
  \BibitemOpen
  \bibfield  {author} {\bibinfo {author} {\bibfnamefont {J.}~\bibnamefont
  {Aalbers}}, \bibinfo {author} {\bibfnamefont {D.~S.}\ \bibnamefont {Akerib}},
  \bibinfo {author} {\bibfnamefont {C.~W.}\ \bibnamefont {Akerlof}}, \bibinfo
  {author} {\bibfnamefont {A.~K.}\ \bibnamefont {Al~Musalhi}}, \bibinfo
  {author} {\bibfnamefont {F.}~\bibnamefont {Alder}}, \bibinfo {author}
  {\bibfnamefont {A.}~\bibnamefont {Alqahtani}}, \bibinfo {author}
  {\bibfnamefont {S.~K.}\ \bibnamefont {Alsum}}, \bibinfo {author}
  {\bibfnamefont {C.~S.}\ \bibnamefont {Amarasinghe}}, \bibinfo {author}
  {\bibfnamefont {A.}~\bibnamefont {Ames}}, \bibinfo {author} {\bibfnamefont
  {T.~J.}\ \bibnamefont {Anderson}}, \emph {et~al.} (\bibinfo {collaboration}
  {LUX-ZEPLIN Collaboration}),\ }\bibfield  {title} {\bibinfo {title} {First
  dark matter search results from the {LUX-ZEPLIN (LZ)} experiment},\ }\href
  {https://doi.org/10.1103/PhysRevLett.131.041002} {\bibfield  {journal}
  {\bibinfo  {journal} {Phys. Rev. Lett.}\ }\textbf {\bibinfo {volume} {131}},\
  \bibinfo {pages} {041002} (\bibinfo {year} {2023})}\BibitemShut {NoStop}%
\bibitem [{\citenamefont {Amerio}\ \emph {et~al.}(2004)\citenamefont {Amerio},
  \citenamefont {Amoruso}, \citenamefont {Antonello}, \citenamefont {Aprili},
  \citenamefont {Armenante}, \citenamefont {Arneodo}, \citenamefont
  {Badertscher}, \citenamefont {Baiboussinov}, \citenamefont {Ceolin},
  \citenamefont {Battistoni} \emph {et~al.}}]{ICARUSDetector}%
  \BibitemOpen
  \bibfield  {author} {\bibinfo {author} {\bibfnamefont {S.}~\bibnamefont
  {Amerio}}, \bibinfo {author} {\bibfnamefont {S.}~\bibnamefont {Amoruso}},
  \bibinfo {author} {\bibfnamefont {M.}~\bibnamefont {Antonello}}, \bibinfo
  {author} {\bibfnamefont {P.}~\bibnamefont {Aprili}}, \bibinfo {author}
  {\bibfnamefont {M.}~\bibnamefont {Armenante}}, \bibinfo {author}
  {\bibfnamefont {F.}~\bibnamefont {Arneodo}}, \bibinfo {author} {\bibfnamefont
  {A.}~\bibnamefont {Badertscher}}, \bibinfo {author} {\bibfnamefont
  {B.}~\bibnamefont {Baiboussinov}}, \bibinfo {author} {\bibfnamefont {M.~B.}\
  \bibnamefont {Ceolin}}, \bibinfo {author} {\bibfnamefont {G.}~\bibnamefont
  {Battistoni}}, \emph {et~al.},\ }\bibfield  {title} {\bibinfo {title}
  {Design, construction and tests of the {ICARUS T600} detector},\ }\href
  {https://doi.org/https://doi.org/10.1016/j.nima.2004.02.044} {\bibfield
  {journal} {\bibinfo  {journal} {Nucl. Instrum. Methods Phys. Res., Sect. A}\
  }\textbf {\bibinfo {volume} {527}},\ \bibinfo {pages} {329} (\bibinfo {year}
  {2004})}\BibitemShut {NoStop}%
\bibitem [{\citenamefont {Acciarri}\ \emph {et~al.}(2017)\citenamefont
  {Acciarri}, \citenamefont {Adams}, \citenamefont {An}, \citenamefont
  {Aparicio}, \citenamefont {Aponte}, \citenamefont {Asaadi}, \citenamefont
  {Auger}, \citenamefont {Ayoub}, \citenamefont {Bagby}, \citenamefont {Baller}
  \emph {et~al.}}]{MicroBooNEDEtector}%
  \BibitemOpen
  \bibfield  {author} {\bibinfo {author} {\bibfnamefont {R.}~\bibnamefont
  {Acciarri}}, \bibinfo {author} {\bibfnamefont {C.}~\bibnamefont {Adams}},
  \bibinfo {author} {\bibfnamefont {R.}~\bibnamefont {An}}, \bibinfo {author}
  {\bibfnamefont {A.}~\bibnamefont {Aparicio}}, \bibinfo {author}
  {\bibfnamefont {S.}~\bibnamefont {Aponte}}, \bibinfo {author} {\bibfnamefont
  {J.}~\bibnamefont {Asaadi}}, \bibinfo {author} {\bibfnamefont
  {M.}~\bibnamefont {Auger}}, \bibinfo {author} {\bibfnamefont
  {N.}~\bibnamefont {Ayoub}}, \bibinfo {author} {\bibfnamefont
  {L.}~\bibnamefont {Bagby}}, \bibinfo {author} {\bibfnamefont
  {B.}~\bibnamefont {Baller}}, \emph {et~al.},\ }\bibfield  {title} {\bibinfo
  {title} {Design and construction of the {MicroBooNE} detector},\ }\href
  {https://doi.org/10.1088/1748-0221/12/02/P02017} {\bibfield  {journal}
  {\bibinfo  {journal} {J. Instrum.}\ }\textbf {\bibinfo {volume} {12}}\bibinfo
   {number} { (02)},\ \bibinfo {pages} {P02017}}\BibitemShut {NoStop}%
\bibitem [{\citenamefont {Abi}\ \emph {et~al.}(2017)\citenamefont {Abi},
  \citenamefont {Acciarri}, \citenamefont {Acero}, \citenamefont {Adamowski},
  \citenamefont {Adams}, \citenamefont {Adams}, \citenamefont {Adamson},
  \citenamefont {Adinolfi}, \citenamefont {Ahmad}, \citenamefont {Albright}
  \emph {et~al.}}]{ProtoDUNEDetector}%
  \BibitemOpen
\bibfield  {number} {  }\bibfield  {author} {\bibinfo {author} {\bibfnamefont
  {B.}~\bibnamefont {Abi}}, \bibinfo {author} {\bibfnamefont {R.}~\bibnamefont
  {Acciarri}}, \bibinfo {author} {\bibfnamefont {M.~A.}\ \bibnamefont {Acero}},
  \bibinfo {author} {\bibfnamefont {M.}~\bibnamefont {Adamowski}}, \bibinfo
  {author} {\bibfnamefont {C.}~\bibnamefont {Adams}}, \bibinfo {author}
  {\bibfnamefont {D.~L.}\ \bibnamefont {Adams}}, \bibinfo {author}
  {\bibfnamefont {P.}~\bibnamefont {Adamson}}, \bibinfo {author} {\bibfnamefont
  {M.}~\bibnamefont {Adinolfi}}, \bibinfo {author} {\bibfnamefont
  {Z.}~\bibnamefont {Ahmad}}, \bibinfo {author} {\bibfnamefont {C.~H.}\
  \bibnamefont {Albright}}, \emph {et~al.},\ }\href@noop {} {\bibinfo {title}
  {The single-phase {ProtoDUNE} technical design report}} (\bibinfo {year}
  {2017}),\ \Eprint {https://arxiv.org/abs/1706.07081} {arXiv:1706.07081
  [physics.ins-det]} \BibitemShut {NoStop}%
\bibitem [{\citenamefont {Abi}\ \emph {et~al.}(2020)\citenamefont {Abi},
  \citenamefont {Acciarri}, \citenamefont {Acero}, \citenamefont {Adamov},
  \citenamefont {Adams}, \citenamefont {Adinolfi}, \citenamefont {Ahmad},
  \citenamefont {Ahmed}, \citenamefont {Alion}, \citenamefont {Monsalve} \emph
  {et~al.}}]{DUNEDetector}%
  \BibitemOpen
  \bibfield  {author} {\bibinfo {author} {\bibfnamefont {B.}~\bibnamefont
  {Abi}}, \bibinfo {author} {\bibfnamefont {R.}~\bibnamefont {Acciarri}},
  \bibinfo {author} {\bibfnamefont {M.~A.}\ \bibnamefont {Acero}}, \bibinfo
  {author} {\bibfnamefont {G.}~\bibnamefont {Adamov}}, \bibinfo {author}
  {\bibfnamefont {D.}~\bibnamefont {Adams}}, \bibinfo {author} {\bibfnamefont
  {M.}~\bibnamefont {Adinolfi}}, \bibinfo {author} {\bibfnamefont
  {Z.}~\bibnamefont {Ahmad}}, \bibinfo {author} {\bibfnamefont
  {J.}~\bibnamefont {Ahmed}}, \bibinfo {author} {\bibfnamefont
  {T.}~\bibnamefont {Alion}}, \bibinfo {author} {\bibfnamefont {S.~A.}\
  \bibnamefont {Monsalve}}, \emph {et~al.},\ }\href@noop {} {} (\bibinfo {year}
  {2020}),\ \Eprint {https://arxiv.org/abs/2002.03005} {2002.03005}
  \BibitemShut {NoStop}%
\bibitem [{\citenamefont {Workman}\ \emph {et~al.}(2022)\citenamefont
  {Workman}, \citenamefont {Burkert}, \citenamefont {Crede}, \citenamefont
  {Klempt}, \citenamefont {Thoma}, \citenamefont {Tiator}, \citenamefont
  {Agashe}, \citenamefont {Aielli}, \citenamefont {Allanach}, \citenamefont
  {Amsler} \emph {et~al.}}]{PDG}%
  \BibitemOpen
  \bibfield  {author} {\bibinfo {author} {\bibfnamefont {R.~L.}\ \bibnamefont
  {Workman}}, \bibinfo {author} {\bibfnamefont {V.}~\bibnamefont {Burkert}},
  \bibinfo {author} {\bibfnamefont {V.}~\bibnamefont {Crede}}, \bibinfo
  {author} {\bibfnamefont {E.}~\bibnamefont {Klempt}}, \bibinfo {author}
  {\bibfnamefont {U.}~\bibnamefont {Thoma}}, \bibinfo {author} {\bibfnamefont
  {L.}~\bibnamefont {Tiator}}, \bibinfo {author} {\bibfnamefont
  {K.}~\bibnamefont {Agashe}}, \bibinfo {author} {\bibfnamefont
  {G.}~\bibnamefont {Aielli}}, \bibinfo {author} {\bibfnamefont
  {B.}~\bibnamefont {Allanach}}, \bibinfo {author} {\bibfnamefont
  {C.}~\bibnamefont {Amsler}}, \emph {et~al.} (\bibinfo {collaboration}
  {Particle Data Group}),\ }\bibfield  {title} {\bibinfo {title} {{Review of
  Particle Physics}},\ }\href {https://doi.org/10.1093/ptep/ptac097} {\bibfield
   {journal} {\bibinfo  {journal} {Prog. Theor. Exp. Phys.}\ }\textbf {\bibinfo
  {volume} {2022}} (\bibinfo {year} {2022})},\ \bibinfo {note}
  {083C01}\BibitemShut {NoStop}%
\bibitem [{\citenamefont {Ambrosi}\ \emph {et~al.}(2017)\citenamefont
  {Ambrosi}, \citenamefont {Choutko}, \citenamefont {Delgado}, \citenamefont
  {Oliva}, \citenamefont {Yan},\ and\ \citenamefont {Li}}]{AMSSiTracker}%
  \BibitemOpen
  \bibfield  {author} {\bibinfo {author} {\bibfnamefont {G.}~\bibnamefont
  {Ambrosi}}, \bibinfo {author} {\bibfnamefont {V.}~\bibnamefont {Choutko}},
  \bibinfo {author} {\bibfnamefont {C.}~\bibnamefont {Delgado}}, \bibinfo
  {author} {\bibfnamefont {A.}~\bibnamefont {Oliva}}, \bibinfo {author}
  {\bibfnamefont {Q.}~\bibnamefont {Yan}},\ and\ \bibinfo {author}
  {\bibfnamefont {Y.}~\bibnamefont {Li}},\ }\bibfield  {title} {\bibinfo
  {title} {The spatial resolution of the silicon tracker of the alpha magnetic
  spectrometer},\ }\href
  {https://doi.org/https://doi.org/10.1016/j.nima.2017.07.014} {\bibfield
  {journal} {\bibinfo  {journal} {Nucl. Instrum. Methods Phys. Res., Sect. A}\
  }\textbf {\bibinfo {volume} {869}},\ \bibinfo {pages} {29} (\bibinfo {year}
  {2017})}\BibitemShut {NoStop}%
\bibitem [{\citenamefont {Straulino}\ \emph {et~al.}(2004)\citenamefont
  {Straulino}, \citenamefont {Adriani}, \citenamefont {Bonechi}, \citenamefont
  {Bongi}, \citenamefont {Castellini}, \citenamefont {D’Alessandro},
  \citenamefont {Gabbanini}, \citenamefont {Grandi}, \citenamefont {Papini},
  \citenamefont {Ricciarini} \emph {et~al.}}]{PAMELASiTracker}%
  \BibitemOpen
  \bibfield  {author} {\bibinfo {author} {\bibfnamefont {S.}~\bibnamefont
  {Straulino}}, \bibinfo {author} {\bibfnamefont {O.}~\bibnamefont {Adriani}},
  \bibinfo {author} {\bibfnamefont {L.}~\bibnamefont {Bonechi}}, \bibinfo
  {author} {\bibfnamefont {M.}~\bibnamefont {Bongi}}, \bibinfo {author}
  {\bibfnamefont {G.}~\bibnamefont {Castellini}}, \bibinfo {author}
  {\bibfnamefont {R.}~\bibnamefont {D’Alessandro}}, \bibinfo {author}
  {\bibfnamefont {A.}~\bibnamefont {Gabbanini}}, \bibinfo {author}
  {\bibfnamefont {M.}~\bibnamefont {Grandi}}, \bibinfo {author} {\bibfnamefont
  {P.}~\bibnamefont {Papini}}, \bibinfo {author} {\bibfnamefont
  {S.}~\bibnamefont {Ricciarini}}, \emph {et~al.},\ }\bibfield  {title}
  {\bibinfo {title} {The {PAMELA} silicon tracker},\ }\href
  {https://doi.org/https://doi.org/10.1016/j.nima.2004.05.067} {\bibfield
  {journal} {\bibinfo  {journal} {Nucl. Instrum. Methods Phys. Res., Sect. A}\
  }\textbf {\bibinfo {volume} {530}},\ \bibinfo {pages} {168} (\bibinfo {year}
  {2004})}\BibitemShut {NoStop}%
\bibitem [{\citenamefont {Chatrchyan}\ \emph {et~al.}(2010)\citenamefont
  {Chatrchyan}, \citenamefont {Khachatryan}, \citenamefont {Sirunyan},
  \citenamefont {Adam}, \citenamefont {Arnold}, \citenamefont {Bergauer},
  \citenamefont {Bergauer}, \citenamefont {Dragicevic}, \citenamefont
  {Eichberger}, \citenamefont {Er\"o} \emph {et~al.}}]{CMSSiTracker}%
  \BibitemOpen
  \bibfield  {author} {\bibinfo {author} {\bibfnamefont {S.}~\bibnamefont
  {Chatrchyan}}, \bibinfo {author} {\bibfnamefont {V.}~\bibnamefont
  {Khachatryan}}, \bibinfo {author} {\bibfnamefont {A.}~\bibnamefont
  {Sirunyan}}, \bibinfo {author} {\bibfnamefont {W.}~\bibnamefont {Adam}},
  \bibinfo {author} {\bibfnamefont {B.}~\bibnamefont {Arnold}}, \bibinfo
  {author} {\bibfnamefont {H.}~\bibnamefont {Bergauer}}, \bibinfo {author}
  {\bibfnamefont {T.}~\bibnamefont {Bergauer}}, \bibinfo {author}
  {\bibfnamefont {M.}~\bibnamefont {Dragicevic}}, \bibinfo {author}
  {\bibfnamefont {M.}~\bibnamefont {Eichberger}}, \bibinfo {author}
  {\bibfnamefont {J.}~\bibnamefont {Er\"o}}, \emph {et~al.} (\bibinfo
  {collaboration} {CMS Collaboration}),\ }\bibfield  {title} {\bibinfo {title}
  {Alignment of the {CMS} silicon tracker during commissioning with cosmic
  rays},\ }\href {https://doi.org/10.1088/1748-0221/5/03/T03009} {\bibfield
  {journal} {\bibinfo  {journal} {J. Instrum.}\ }\textbf {\bibinfo {volume}
  {5}}\bibinfo  {number} { (03)},\ \bibinfo {pages} {T03009}}\BibitemShut
  {NoStop}%
\bibitem [{\citenamefont {Leboucher}\ \emph {et~al.}(2022)\citenamefont
  {Leboucher}, \citenamefont {Adamczyk}, \citenamefont {Aggarwal},
  \citenamefont {Aihara}, \citenamefont {Aziz}, \citenamefont {Bacher},
  \citenamefont {Bahinipati}, \citenamefont {Batignani}, \citenamefont
  {Baudot}, \citenamefont {Behera} \emph {et~al.}}]{BelleIISiTracker}%
  \BibitemOpen
\bibfield  {number} {  }\bibfield  {author} {\bibinfo {author} {\bibfnamefont
  {R.}~\bibnamefont {Leboucher}}, \bibinfo {author} {\bibfnamefont
  {K.}~\bibnamefont {Adamczyk}}, \bibinfo {author} {\bibfnamefont
  {L.}~\bibnamefont {Aggarwal}}, \bibinfo {author} {\bibfnamefont
  {H.}~\bibnamefont {Aihara}}, \bibinfo {author} {\bibfnamefont
  {T.}~\bibnamefont {Aziz}}, \bibinfo {author} {\bibfnamefont {S.}~\bibnamefont
  {Bacher}}, \bibinfo {author} {\bibfnamefont {S.}~\bibnamefont {Bahinipati}},
  \bibinfo {author} {\bibfnamefont {G.}~\bibnamefont {Batignani}}, \bibinfo
  {author} {\bibfnamefont {J.}~\bibnamefont {Baudot}}, \bibinfo {author}
  {\bibfnamefont {P.}~\bibnamefont {Behera}}, \emph {et~al.} (\bibinfo
  {collaboration} {Belle-II SVC Collaboration}),\ }\bibfield  {title} {\bibinfo
  {title} {Measurement of the cluster position resolution of the belle ii
  silicon vertex detector},\ }\href
  {https://doi.org/https://doi.org/10.1016/j.nima.2022.166746} {\bibfield
  {journal} {\bibinfo  {journal} {Nucl. Instrum. Methods Phys. Res., Sect. A}\
  }\textbf {\bibinfo {volume} {1033}},\ \bibinfo {pages} {166746} (\bibinfo
  {year} {2022})}\BibitemShut {NoStop}%
\bibitem [{\citenamefont {Lange}\ \emph {et~al.}(2016)\citenamefont {Lange},
  \citenamefont {Adamczyk}, \citenamefont {Avoni}, \citenamefont {Banas},
  \citenamefont {Brandt}, \citenamefont {Bruschi}, \citenamefont {Buglewicz},
  \citenamefont {Cavallaro}, \citenamefont {Caforio}, \citenamefont {Chiodini}
  \emph {et~al.}}]{ATLASSiTracker}%
  \BibitemOpen
  \bibfield  {author} {\bibinfo {author} {\bibfnamefont {J.}~\bibnamefont
  {Lange}}, \bibinfo {author} {\bibfnamefont {L.}~\bibnamefont {Adamczyk}},
  \bibinfo {author} {\bibfnamefont {G.}~\bibnamefont {Avoni}}, \bibinfo
  {author} {\bibfnamefont {E.}~\bibnamefont {Banas}}, \bibinfo {author}
  {\bibfnamefont {A.}~\bibnamefont {Brandt}}, \bibinfo {author} {\bibfnamefont
  {M.}~\bibnamefont {Bruschi}}, \bibinfo {author} {\bibfnamefont
  {P.}~\bibnamefont {Buglewicz}}, \bibinfo {author} {\bibfnamefont
  {E.}~\bibnamefont {Cavallaro}}, \bibinfo {author} {\bibfnamefont
  {D.}~\bibnamefont {Caforio}}, \bibinfo {author} {\bibfnamefont
  {G.}~\bibnamefont {Chiodini}}, \emph {et~al.},\ }\bibfield  {title} {\bibinfo
  {title} {Beam tests of an integrated prototype of the {ATLAS} forward proton
  detector},\ }\href {https://doi.org/10.1088/1748-0221/11/09/P09005}
  {\bibfield  {journal} {\bibinfo  {journal} {J. Instrum.}\ }\textbf {\bibinfo
  {volume} {11}}\bibinfo  {number} { (09)},\ \bibinfo {pages}
  {P09005}}\BibitemShut {NoStop}%
\bibitem [{\citenamefont {Hartmann}(2012)}]{SiTrackerReview}%
  \BibitemOpen
\bibfield  {number} {  }\bibfield  {author} {\bibinfo {author} {\bibfnamefont
  {F.}~\bibnamefont {Hartmann}},\ }\bibfield  {title} {\bibinfo {title}
  {Silicon tracking detectors in high-energy physics},\ }\href
  {https://doi.org/https://doi.org/10.1016/j.nima.2011.11.005} {\bibfield
  {journal} {\bibinfo  {journal} {Nucl. Instrum. Methods Phys. Res., Sect. A}\
  }\textbf {\bibinfo {volume} {666}},\ \bibinfo {pages} {25} (\bibinfo {year}
  {2012})}\BibitemShut {NoStop}%
\bibitem [{\citenamefont {Ye}\ \emph {et~al.}(2021)\citenamefont {Ye},
  \citenamefont {Abudinen}, \citenamefont {Ackermann}, \citenamefont {Ahlburg},
  \citenamefont {Albalawi}, \citenamefont {Alonso}, \citenamefont {Andricek},
  \citenamefont {Ayad}, \citenamefont {Babu}, \citenamefont {Bilka} \emph
  {et~al.}}]{BelleIIDEPFET}%
  \BibitemOpen
  \bibfield  {author} {\bibinfo {author} {\bibfnamefont {H.}~\bibnamefont
  {Ye}}, \bibinfo {author} {\bibfnamefont {F.}~\bibnamefont {Abudinen}},
  \bibinfo {author} {\bibfnamefont {K.}~\bibnamefont {Ackermann}}, \bibinfo
  {author} {\bibfnamefont {P.}~\bibnamefont {Ahlburg}}, \bibinfo {author}
  {\bibfnamefont {M.}~\bibnamefont {Albalawi}}, \bibinfo {author}
  {\bibfnamefont {O.}~\bibnamefont {Alonso}}, \bibinfo {author} {\bibfnamefont
  {L.}~\bibnamefont {Andricek}}, \bibinfo {author} {\bibfnamefont
  {R.}~\bibnamefont {Ayad}}, \bibinfo {author} {\bibfnamefont {V.}~\bibnamefont
  {Babu}}, \bibinfo {author} {\bibfnamefont {T.}~\bibnamefont {Bilka}}, \emph
  {et~al.},\ }\bibfield  {title} {\bibinfo {title} {Commissioning and
  performance of the belle ii pixel detector},\ }\href
  {https://doi.org/https://doi.org/10.1016/j.nima.2020.164875} {\bibfield
  {journal} {\bibinfo  {journal} {Nucl. Instrum. Methods Phys. Res., Sect. A}\
  }\textbf {\bibinfo {volume} {987}},\ \bibinfo {pages} {164875} (\bibinfo
  {year} {2021})}\BibitemShut {NoStop}%
\bibitem [{\citenamefont {{Aglieri Rinella}}(2017)}]{ALICEMAPS}%
  \BibitemOpen
  \bibfield  {author} {\bibinfo {author} {\bibfnamefont {G.}~\bibnamefont
  {{Aglieri Rinella}}},\ }\bibfield  {title} {\bibinfo {title} {The {ALPIDE}
  pixel sensor chip for the upgrade of the {ALICE} inner tracking system},\
  }\href {https://doi.org/https://doi.org/10.1016/j.nima.2016.05.016}
  {\bibfield  {journal} {\bibinfo  {journal} {Nucl. Instrum. Methods Phys.
  Res., Sect. A}\ }\textbf {\bibinfo {volume} {845}},\ \bibinfo {pages} {583}
  (\bibinfo {year} {2017})}\BibitemShut {NoStop}%
\bibitem [{\citenamefont {Adam}\ \emph {et~al.}(2021)\citenamefont {Adam},
  \citenamefont {Bergauer}, \citenamefont {Blöch}, \citenamefont {Dragicevic},
  \citenamefont {Frühwirth}, \citenamefont {Hinger}, \citenamefont
  {Steininger}, \citenamefont {Beaumont}, \citenamefont {Croce}, \citenamefont
  {Janssen} \emph {et~al.}}]{CMSPixel}%
  \BibitemOpen
  \bibfield  {author} {\bibinfo {author} {\bibfnamefont {W.}~\bibnamefont
  {Adam}}, \bibinfo {author} {\bibfnamefont {T.}~\bibnamefont {Bergauer}},
  \bibinfo {author} {\bibfnamefont {D.}~\bibnamefont {Blöch}}, \bibinfo
  {author} {\bibfnamefont {M.}~\bibnamefont {Dragicevic}}, \bibinfo {author}
  {\bibfnamefont {R.}~\bibnamefont {Frühwirth}}, \bibinfo {author}
  {\bibfnamefont {V.}~\bibnamefont {Hinger}}, \bibinfo {author} {\bibfnamefont
  {H.}~\bibnamefont {Steininger}}, \bibinfo {author} {\bibfnamefont
  {W.}~\bibnamefont {Beaumont}}, \bibinfo {author} {\bibfnamefont {D.~D.}\
  \bibnamefont {Croce}}, \bibinfo {author} {\bibfnamefont {X.}~\bibnamefont
  {Janssen}}, \emph {et~al.},\ }\bibfield  {title} {\bibinfo {title} {The {CMS}
  phase-1 pixel detector upgrade},\ }\href
  {https://doi.org/10.1088/1748-0221/16/02/P02027} {\bibfield  {journal}
  {\bibinfo  {journal} {J. Instrum.}\ }\textbf {\bibinfo {volume} {16}}\bibinfo
   {number} { (02)},\ \bibinfo {pages} {P02027}}\BibitemShut {NoStop}%
\bibitem [{\citenamefont {Peric}(2012)}]{ATLASPixel}%
  \BibitemOpen
\bibfield  {number} {  }\bibfield  {author} {\bibinfo {author} {\bibfnamefont
  {I.}~\bibnamefont {Peric}},\ }\bibfield  {title} {\bibinfo {title} {Active
  pixel sensors in high-voltage {CMOS} technologies for {ATLAS}},\ }\href
  {https://doi.org/10.1088/1748-0221/7/08/C08002} {\bibfield  {journal}
  {\bibinfo  {journal} {J. Instrum.}\ }\textbf {\bibinfo {volume} {7}}\bibinfo
  {number} { (08)},\ \bibinfo {pages} {C08002}}\BibitemShut {NoStop}%
\bibitem [{\citenamefont {Buchanan}(2017)}]{LHCbPixel}%
  \BibitemOpen
\bibfield  {number} {  }\bibfield  {author} {\bibinfo {author} {\bibfnamefont
  {E.}~\bibnamefont {Buchanan}},\ }\bibfield  {title} {\bibinfo {title} {The
  {LHCb} vertex locator ({VELO}) pixel detector upgrade},\ }\href
  {https://doi.org/10.1088/1748-0221/12/01/C01013} {\bibfield  {journal}
  {\bibinfo  {journal} {J. Instrum.}\ }\textbf {\bibinfo {volume} {12}}\bibinfo
   {number} { (01)},\ \bibinfo {pages} {C01013}}\BibitemShut {NoStop}%
\bibitem [{\citenamefont {{De Serio}}\ \emph {et~al.}(2005)\citenamefont {{De
  Serio}}, \citenamefont {Ieva}, \citenamefont {Muciaccia}, \citenamefont
  {Simone}, \citenamefont {Cozzi}, \citenamefont {Giacomelli}, \citenamefont
  {Patrizii}, \citenamefont {Sirri}, \citenamefont {Blokhin}, \citenamefont
  {Buontempo} \emph {et~al.}}]{NIMA.554.247}%
  \BibitemOpen
\bibfield  {number} {  }\bibfield  {author} {\bibinfo {author} {\bibfnamefont
  {M.}~\bibnamefont {{De Serio}}}, \bibinfo {author} {\bibfnamefont
  {M.}~\bibnamefont {Ieva}}, \bibinfo {author} {\bibfnamefont {M.}~\bibnamefont
  {Muciaccia}}, \bibinfo {author} {\bibfnamefont {S.}~\bibnamefont {Simone}},
  \bibinfo {author} {\bibfnamefont {M.}~\bibnamefont {Cozzi}}, \bibinfo
  {author} {\bibfnamefont {G.}~\bibnamefont {Giacomelli}}, \bibinfo {author}
  {\bibfnamefont {L.}~\bibnamefont {Patrizii}}, \bibinfo {author}
  {\bibfnamefont {G.}~\bibnamefont {Sirri}}, \bibinfo {author} {\bibfnamefont
  {S.}~\bibnamefont {Blokhin}}, \bibinfo {author} {\bibfnamefont
  {S.}~\bibnamefont {Buontempo}}, \emph {et~al.},\ }\bibfield  {title}
  {\bibinfo {title} {High precision measurements with nuclear emulsions using
  fast automated microscopes},\ }\href
  {https://doi.org/https://doi.org/10.1016/j.nima.2005.08.017} {\bibfield
  {journal} {\bibinfo  {journal} {Nucl. Instrum. Methods Phys. Res., Sect. A}\
  }\textbf {\bibinfo {volume} {554}},\ \bibinfo {pages} {247} (\bibinfo {year}
  {2005})}\BibitemShut {NoStop}%
\bibitem [{\citenamefont {Naka}\ \emph {et~al.}(2013)\citenamefont {Naka},
  \citenamefont {Asada}, \citenamefont {Katsuragawa}, \citenamefont {Hakamata},
  \citenamefont {Yoshimoto}, \citenamefont {Kuwabara}, \citenamefont
  {Nakamura}, \citenamefont {Sato}, \citenamefont {Nakano}, \citenamefont
  {Tawara} \emph {et~al.}}]{NIMA.718.519}%
  \BibitemOpen
  \bibfield  {author} {\bibinfo {author} {\bibfnamefont {T.}~\bibnamefont
  {Naka}}, \bibinfo {author} {\bibfnamefont {T.}~\bibnamefont {Asada}},
  \bibinfo {author} {\bibfnamefont {T.}~\bibnamefont {Katsuragawa}}, \bibinfo
  {author} {\bibfnamefont {K.}~\bibnamefont {Hakamata}}, \bibinfo {author}
  {\bibfnamefont {M.}~\bibnamefont {Yoshimoto}}, \bibinfo {author}
  {\bibfnamefont {K.}~\bibnamefont {Kuwabara}}, \bibinfo {author}
  {\bibfnamefont {M.}~\bibnamefont {Nakamura}}, \bibinfo {author}
  {\bibfnamefont {O.}~\bibnamefont {Sato}}, \bibinfo {author} {\bibfnamefont
  {T.}~\bibnamefont {Nakano}}, \bibinfo {author} {\bibfnamefont
  {Y.}~\bibnamefont {Tawara}}, \emph {et~al.},\ }\bibfield  {title} {\bibinfo
  {title} {Fine grained nuclear emulsion for higher resolution tracking
  detector},\ }\href
  {https://doi.org/https://doi.org/10.1016/j.nima.2012.11.106} {\bibfield
  {journal} {\bibinfo  {journal} {Nucl. Instrum. Methods Phys. Res., Sect. A}\
  }\textbf {\bibinfo {volume} {718}},\ \bibinfo {pages} {519} (\bibinfo {year}
  {2013})}\BibitemShut {NoStop}%
\bibitem [{\citenamefont {Nakamura}\ \emph {et~al.}(2006)\citenamefont
  {Nakamura}, \citenamefont {Ariga}, \citenamefont {Ban}, \citenamefont
  {Fukuda}, \citenamefont {Fukuda}, \citenamefont {Fujioka}, \citenamefont
  {Furukawa}, \citenamefont {Hamada}, \citenamefont {Hayashi}, \citenamefont
  {Hiramatsu} \emph {et~al.}}]{NIMA.556.80}%
  \BibitemOpen
  \bibfield  {author} {\bibinfo {author} {\bibfnamefont {T.}~\bibnamefont
  {Nakamura}}, \bibinfo {author} {\bibfnamefont {A.}~\bibnamefont {Ariga}},
  \bibinfo {author} {\bibfnamefont {T.}~\bibnamefont {Ban}}, \bibinfo {author}
  {\bibfnamefont {T.}~\bibnamefont {Fukuda}}, \bibinfo {author} {\bibfnamefont
  {T.}~\bibnamefont {Fukuda}}, \bibinfo {author} {\bibfnamefont
  {T.}~\bibnamefont {Fujioka}}, \bibinfo {author} {\bibfnamefont
  {T.}~\bibnamefont {Furukawa}}, \bibinfo {author} {\bibfnamefont
  {K.}~\bibnamefont {Hamada}}, \bibinfo {author} {\bibfnamefont
  {H.}~\bibnamefont {Hayashi}}, \bibinfo {author} {\bibfnamefont
  {S.}~\bibnamefont {Hiramatsu}}, \emph {et~al.},\ }\bibfield  {title}
  {\bibinfo {title} {The {OPERA} film: New nuclear emulsion for large-scale,
  high-precision experiments},\ }\href
  {https://doi.org/https://doi.org/10.1016/j.nima.2005.08.109} {\bibfield
  {journal} {\bibinfo  {journal} {Nucl. Instrum. Methods Phys. Res., Sect. A}\
  }\textbf {\bibinfo {volume} {556}},\ \bibinfo {pages} {80} (\bibinfo {year}
  {2006})}\BibitemShut {NoStop}%
\bibitem [{\citenamefont {DeStefano}\ \emph {et~al.}()\citenamefont
  {DeStefano}, \citenamefont {Pegahan}, \citenamefont {Novikova}, \citenamefont
  {Mikhailov}, \citenamefont {Aubin}, \citenamefont {Averett}, \citenamefont
  {Zhang}, \citenamefont {Park}, \citenamefont {Camsonne}, \citenamefont
  {Ramaswamy},\ and\ \citenamefont {Malinovskaya}}]{DAMOPZeeman}%
  \BibitemOpen
  \bibfield  {author} {\bibinfo {author} {\bibfnamefont {N.~C.}\ \bibnamefont
  {DeStefano}}, \bibinfo {author} {\bibfnamefont {S.}~\bibnamefont {Pegahan}},
  \bibinfo {author} {\bibfnamefont {I.~B.}\ \bibnamefont {Novikova}}, \bibinfo
  {author} {\bibfnamefont {E.~E.}\ \bibnamefont {Mikhailov}}, \bibinfo {author}
  {\bibfnamefont {S.}~\bibnamefont {Aubin}}, \bibinfo {author} {\bibfnamefont
  {T.~D.}\ \bibnamefont {Averett}}, \bibinfo {author} {\bibfnamefont
  {S.}~\bibnamefont {Zhang}}, \bibinfo {author} {\bibfnamefont
  {G.}~\bibnamefont {Park}}, \bibinfo {author} {\bibfnamefont {A.}~\bibnamefont
  {Camsonne}}, \bibinfo {author} {\bibfnamefont {A.}~\bibnamefont
  {Ramaswamy}},\ and\ \bibinfo {author} {\bibfnamefont {S.~A.}\ \bibnamefont
  {Malinovskaya}},\ }\bibfield  {title} {\bibinfo {title} {Development of a
  rydberg atom-based apparatus for tracking charged particles},\ }\bibinfo
  {note} {development of a Rydberg Atom-Based Apparatus for Tracking Charged
  Particles (54th Annual Meeting of the APS Division of Atomic, Molecular and
  Optical Physics, Spokane, WA, 2023)}\BibitemShut {NoStop}%
\bibitem [{\citenamefont {Aubin}\ \emph {et~al.}()\citenamefont {Aubin},
  \citenamefont {Pegahan}, \citenamefont {Averett}, \citenamefont {Mikhailov},
  \citenamefont {Novikova}, \citenamefont {DeStefano}, \citenamefont {Zhang},
  \citenamefont {Camsonne}, \citenamefont {Park},\ and\ \citenamefont
  {Ramaswamy}}]{DAMOPStark}%
  \BibitemOpen
  \bibfield  {author} {\bibinfo {author} {\bibfnamefont {S.}~\bibnamefont
  {Aubin}}, \bibinfo {author} {\bibfnamefont {S.}~\bibnamefont {Pegahan}},
  \bibinfo {author} {\bibfnamefont {T.~D.}\ \bibnamefont {Averett}}, \bibinfo
  {author} {\bibfnamefont {E.~E.}\ \bibnamefont {Mikhailov}}, \bibinfo {author}
  {\bibfnamefont {I.~B.}\ \bibnamefont {Novikova}}, \bibinfo {author}
  {\bibfnamefont {N.~C.}\ \bibnamefont {DeStefano}}, \bibinfo {author}
  {\bibfnamefont {S.}~\bibnamefont {Zhang}}, \bibinfo {author} {\bibfnamefont
  {A.}~\bibnamefont {Camsonne}}, \bibinfo {author} {\bibfnamefont
  {G.}~\bibnamefont {Park}},\ and\ \bibinfo {author} {\bibfnamefont
  {A.}~\bibnamefont {Ramaswamy}},\ }\bibfield  {title} {\bibinfo {title}
  {Imaging charged particle beams with atomic magnetometry},\ }\bibinfo {note}
  {imaging Charged Particle Beams With Atomic Magnetometry (54th Annual Meeting
  of the APS Division of Atomic, Molecular and Optical Physics, Spokane, WA,
  2023)}\BibitemShut {NoStop}%
\bibitem [{\citenamefont {Bernien}\ \emph {et~al.}(2017)\citenamefont
  {Bernien}, \citenamefont {Schwartz}, \citenamefont {Keesling}, \citenamefont
  {Levine}, \citenamefont {Omran}, \citenamefont {Pichler}, \citenamefont
  {Choi}, \citenamefont {Zibrov}, \citenamefont {Endres}, \citenamefont
  {Greiner} \emph {et~al.}}]{Nature.551.579}%
  \BibitemOpen
  \bibfield  {author} {\bibinfo {author} {\bibfnamefont {H.}~\bibnamefont
  {Bernien}}, \bibinfo {author} {\bibfnamefont {S.}~\bibnamefont {Schwartz}},
  \bibinfo {author} {\bibfnamefont {A.}~\bibnamefont {Keesling}}, \bibinfo
  {author} {\bibfnamefont {H.}~\bibnamefont {Levine}}, \bibinfo {author}
  {\bibfnamefont {A.}~\bibnamefont {Omran}}, \bibinfo {author} {\bibfnamefont
  {H.}~\bibnamefont {Pichler}}, \bibinfo {author} {\bibfnamefont
  {S.}~\bibnamefont {Choi}}, \bibinfo {author} {\bibfnamefont {A.~S.}\
  \bibnamefont {Zibrov}}, \bibinfo {author} {\bibfnamefont {M.}~\bibnamefont
  {Endres}}, \bibinfo {author} {\bibfnamefont {M.}~\bibnamefont {Greiner}},
  \emph {et~al.},\ }\bibfield  {title} {\bibinfo {title} {Probing many-body
  dynamics on a 51-atom quantum simulator},\ }\href
  {https://doi.org/10.1038/nature24622} {\bibfield  {journal} {\bibinfo
  {journal} {Nature}\ }\textbf {\bibinfo {volume} {551}},\ \bibinfo {pages}
  {579} (\bibinfo {year} {2017})}\BibitemShut {NoStop}%
\bibitem [{\citenamefont {Cooper}\ \emph {et~al.}(2018)\citenamefont {Cooper},
  \citenamefont {Covey}, \citenamefont {Madjarov}, \citenamefont {Porsev},
  \citenamefont {Safronova},\ and\ \citenamefont {Endres}}]{PhysRevX.8.041055}%
  \BibitemOpen
  \bibfield  {author} {\bibinfo {author} {\bibfnamefont {A.}~\bibnamefont
  {Cooper}}, \bibinfo {author} {\bibfnamefont {J.~P.}\ \bibnamefont {Covey}},
  \bibinfo {author} {\bibfnamefont {I.~S.}\ \bibnamefont {Madjarov}}, \bibinfo
  {author} {\bibfnamefont {S.~G.}\ \bibnamefont {Porsev}}, \bibinfo {author}
  {\bibfnamefont {M.~S.}\ \bibnamefont {Safronova}},\ and\ \bibinfo {author}
  {\bibfnamefont {M.}~\bibnamefont {Endres}},\ }\bibfield  {title} {\bibinfo
  {title} {Alkaline-earth atoms in optical tweezers},\ }\href
  {https://doi.org/10.1103/PhysRevX.8.041055} {\bibfield  {journal} {\bibinfo
  {journal} {Phys. Rev. X}\ }\textbf {\bibinfo {volume} {8}},\ \bibinfo {pages}
  {041055} (\bibinfo {year} {2018})}\BibitemShut {NoStop}%
\bibitem [{\citenamefont {Norcia}\ \emph {et~al.}(2018)\citenamefont {Norcia},
  \citenamefont {Young},\ and\ \citenamefont {Kaufman}}]{PhysRevX.8.041054}%
  \BibitemOpen
  \bibfield  {author} {\bibinfo {author} {\bibfnamefont {M.~A.}\ \bibnamefont
  {Norcia}}, \bibinfo {author} {\bibfnamefont {A.~W.}\ \bibnamefont {Young}},\
  and\ \bibinfo {author} {\bibfnamefont {A.~M.}\ \bibnamefont {Kaufman}},\
  }\bibfield  {title} {\bibinfo {title} {Microscopic control and detection of
  ultracold strontium in optical-tweezer arrays},\ }\href
  {https://doi.org/10.1103/PhysRevX.8.041054} {\bibfield  {journal} {\bibinfo
  {journal} {Phys. Rev. X}\ }\textbf {\bibinfo {volume} {8}},\ \bibinfo {pages}
  {041054} (\bibinfo {year} {2018})}\BibitemShut {NoStop}%
\bibitem [{\citenamefont {Saskin}\ \emph {et~al.}(2019)\citenamefont {Saskin},
  \citenamefont {Wilson}, \citenamefont {Grinkemeyer},\ and\ \citenamefont
  {Thompson}}]{PhysRevLett.122.143002}%
  \BibitemOpen
  \bibfield  {author} {\bibinfo {author} {\bibfnamefont {S.}~\bibnamefont
  {Saskin}}, \bibinfo {author} {\bibfnamefont {J.~T.}\ \bibnamefont {Wilson}},
  \bibinfo {author} {\bibfnamefont {B.}~\bibnamefont {Grinkemeyer}},\ and\
  \bibinfo {author} {\bibfnamefont {J.~D.}\ \bibnamefont {Thompson}},\
  }\bibfield  {title} {\bibinfo {title} {Narrow-line cooling and imaging of
  ytterbium atoms in an optical tweezer array},\ }\href
  {https://doi.org/10.1103/PhysRevLett.122.143002} {\bibfield  {journal}
  {\bibinfo  {journal} {Phys. Rev. Lett.}\ }\textbf {\bibinfo {volume} {122}},\
  \bibinfo {pages} {143002} (\bibinfo {year} {2019})}\BibitemShut {NoStop}%
\bibitem [{\citenamefont {Barredo}\ \emph {et~al.}(2016)\citenamefont
  {Barredo}, \citenamefont {de~Léséleuc}, \citenamefont {Lienhard},
  \citenamefont {Lahaye},\ and\ \citenamefont {Browaeys}}]{Science.354.1021}%
  \BibitemOpen
  \bibfield  {author} {\bibinfo {author} {\bibfnamefont {D.}~\bibnamefont
  {Barredo}}, \bibinfo {author} {\bibfnamefont {S.}~\bibnamefont
  {de~Léséleuc}}, \bibinfo {author} {\bibfnamefont {V.}~\bibnamefont
  {Lienhard}}, \bibinfo {author} {\bibfnamefont {T.}~\bibnamefont {Lahaye}},\
  and\ \bibinfo {author} {\bibfnamefont {A.}~\bibnamefont {Browaeys}},\
  }\bibfield  {title} {\bibinfo {title} {An atom-by-atom assembler of
  defect-free arbitrary two-dimensional atomic arrays},\ }\href
  {https://doi.org/10.1126/science.aah3778} {\bibfield  {journal} {\bibinfo
  {journal} {Science}\ }\textbf {\bibinfo {volume} {354}},\ \bibinfo {pages}
  {1021} (\bibinfo {year} {2016})}\BibitemShut {NoStop}%
\bibitem [{\citenamefont {Kim}\ \emph {et~al.}(2019)\citenamefont {Kim},
  \citenamefont {Kim}, \citenamefont {Lee},\ and\ \citenamefont
  {Ahn}}]{OptExpress.27.2184}%
  \BibitemOpen
  \bibfield  {author} {\bibinfo {author} {\bibfnamefont {H.}~\bibnamefont
  {Kim}}, \bibinfo {author} {\bibfnamefont {M.}~\bibnamefont {Kim}}, \bibinfo
  {author} {\bibfnamefont {W.}~\bibnamefont {Lee}},\ and\ \bibinfo {author}
  {\bibfnamefont {J.}~\bibnamefont {Ahn}},\ }\bibfield  {title} {\bibinfo
  {title} {Gerchberg-saxton algorithm for fast and efficient atom rearrangement
  in optical tweezer traps},\ }\href {https://doi.org/10.1364/OE.27.002184}
  {\bibfield  {journal} {\bibinfo  {journal} {Opt. Express}\ }\textbf {\bibinfo
  {volume} {27}},\ \bibinfo {pages} {2184} (\bibinfo {year}
  {2019})}\BibitemShut {NoStop}%
\bibitem [{\citenamefont {Schymik}\ \emph {et~al.}(2020)\citenamefont
  {Schymik}, \citenamefont {Lienhard}, \citenamefont {Barredo}, \citenamefont
  {Scholl}, \citenamefont {Williams}, \citenamefont {Browaeys},\ and\
  \citenamefont {Lahaye}}]{PhysRevA.102.063107}%
  \BibitemOpen
  \bibfield  {author} {\bibinfo {author} {\bibfnamefont {K.-N.}\ \bibnamefont
  {Schymik}}, \bibinfo {author} {\bibfnamefont {V.}~\bibnamefont {Lienhard}},
  \bibinfo {author} {\bibfnamefont {D.}~\bibnamefont {Barredo}}, \bibinfo
  {author} {\bibfnamefont {P.}~\bibnamefont {Scholl}}, \bibinfo {author}
  {\bibfnamefont {H.}~\bibnamefont {Williams}}, \bibinfo {author}
  {\bibfnamefont {A.}~\bibnamefont {Browaeys}},\ and\ \bibinfo {author}
  {\bibfnamefont {T.}~\bibnamefont {Lahaye}},\ }\bibfield  {title} {\bibinfo
  {title} {Enhanced atom-by-atom assembly of arbitrary tweezer arrays},\ }\href
  {https://doi.org/10.1103/PhysRevA.102.063107} {\bibfield  {journal} {\bibinfo
   {journal} {Phys. Rev. A}\ }\textbf {\bibinfo {volume} {102}},\ \bibinfo
  {pages} {063107} (\bibinfo {year} {2020})}\BibitemShut {NoStop}%
\bibitem [{\citenamefont {Tian}\ \emph {et~al.}(2023)\citenamefont {Tian},
  \citenamefont {Wee}, \citenamefont {Qu}, \citenamefont {Lim}, \citenamefont
  {Datla}, \citenamefont {Koh},\ and\ \citenamefont
  {Loh}}]{PhysRevApplied.19.034048}%
  \BibitemOpen
  \bibfield  {author} {\bibinfo {author} {\bibfnamefont {W.}~\bibnamefont
  {Tian}}, \bibinfo {author} {\bibfnamefont {W.~J.}\ \bibnamefont {Wee}},
  \bibinfo {author} {\bibfnamefont {A.}~\bibnamefont {Qu}}, \bibinfo {author}
  {\bibfnamefont {B.~J.~M.}\ \bibnamefont {Lim}}, \bibinfo {author}
  {\bibfnamefont {P.~R.}\ \bibnamefont {Datla}}, \bibinfo {author}
  {\bibfnamefont {V.~P.~W.}\ \bibnamefont {Koh}},\ and\ \bibinfo {author}
  {\bibfnamefont {H.}~\bibnamefont {Loh}},\ }\bibfield  {title} {\bibinfo
  {title} {Parallel assembly of arbitrary defect-free atom arrays with a
  multitweezer algorithm},\ }\href
  {https://doi.org/10.1103/PhysRevApplied.19.034048} {\bibfield  {journal}
  {\bibinfo  {journal} {Phys. Rev. Appl.}\ }\textbf {\bibinfo {volume} {19}},\
  \bibinfo {pages} {034048} (\bibinfo {year} {2023})}\BibitemShut {NoStop}%
\bibitem [{\citenamefont {Bakr}\ \emph {et~al.}(2009)\citenamefont {Bakr},
  \citenamefont {Gillen}, \citenamefont {Peng}, \citenamefont {F{\"o}lling},\
  and\ \citenamefont {Greiner}}]{Nature.462.74}%
  \BibitemOpen
  \bibfield  {author} {\bibinfo {author} {\bibfnamefont {W.~S.}\ \bibnamefont
  {Bakr}}, \bibinfo {author} {\bibfnamefont {J.~I.}\ \bibnamefont {Gillen}},
  \bibinfo {author} {\bibfnamefont {A.}~\bibnamefont {Peng}}, \bibinfo {author}
  {\bibfnamefont {S.}~\bibnamefont {F{\"o}lling}},\ and\ \bibinfo {author}
  {\bibfnamefont {M.}~\bibnamefont {Greiner}},\ }\bibfield  {title} {\bibinfo
  {title} {A quantum gas microscope for detecting single atoms in a
  hubbard-regime optical lattice},\ }\href
  {https://doi.org/10.1038/nature08482} {\bibfield  {journal} {\bibinfo
  {journal} {Nature}\ }\textbf {\bibinfo {volume} {462}},\ \bibinfo {pages}
  {74} (\bibinfo {year} {2009})}\BibitemShut {NoStop}%
\bibitem [{\citenamefont {Sherson}\ \emph {et~al.}(2010)\citenamefont
  {Sherson}, \citenamefont {Weitenberg}, \citenamefont {Endres}, \citenamefont
  {Cheneau}, \citenamefont {Bloch},\ and\ \citenamefont
  {Kuhr}}]{Nature.467.68}%
  \BibitemOpen
  \bibfield  {author} {\bibinfo {author} {\bibfnamefont {J.~F.}\ \bibnamefont
  {Sherson}}, \bibinfo {author} {\bibfnamefont {C.}~\bibnamefont {Weitenberg}},
  \bibinfo {author} {\bibfnamefont {M.}~\bibnamefont {Endres}}, \bibinfo
  {author} {\bibfnamefont {M.}~\bibnamefont {Cheneau}}, \bibinfo {author}
  {\bibfnamefont {I.}~\bibnamefont {Bloch}},\ and\ \bibinfo {author}
  {\bibfnamefont {S.}~\bibnamefont {Kuhr}},\ }\bibfield  {title} {\bibinfo
  {title} {Single-atom-resolved fluorescence imaging of an atomic mott
  insulator},\ }\href {https://doi.org/10.1038/nature09378} {\bibfield
  {journal} {\bibinfo  {journal} {Nature}\ }\textbf {\bibinfo {volume} {467}},\
  \bibinfo {pages} {68} (\bibinfo {year} {2010})}\BibitemShut {NoStop}%
\bibitem [{\citenamefont {Marti}\ \emph {et~al.}(2018)\citenamefont {Marti},
  \citenamefont {Hutson}, \citenamefont {Goban}, \citenamefont {Campbell},
  \citenamefont {Poli},\ and\ \citenamefont {Ye}}]{PhysRevLett.120.103201}%
  \BibitemOpen
  \bibfield  {author} {\bibinfo {author} {\bibfnamefont {G.~E.}\ \bibnamefont
  {Marti}}, \bibinfo {author} {\bibfnamefont {R.~B.}\ \bibnamefont {Hutson}},
  \bibinfo {author} {\bibfnamefont {A.}~\bibnamefont {Goban}}, \bibinfo
  {author} {\bibfnamefont {S.~L.}\ \bibnamefont {Campbell}}, \bibinfo {author}
  {\bibfnamefont {N.}~\bibnamefont {Poli}},\ and\ \bibinfo {author}
  {\bibfnamefont {J.}~\bibnamefont {Ye}},\ }\bibfield  {title} {\bibinfo
  {title} {Imaging optical frequencies with $100\text{ }\text{
  }\ensuremath{\mu}\mathrm{Hz}$ precision and $1.1\text{ }\text{
  }\ensuremath{\mu}\mathrm{m}$ resolution},\ }\href
  {https://doi.org/10.1103/PhysRevLett.120.103201} {\bibfield  {journal}
  {\bibinfo  {journal} {Phys. Rev. Lett.}\ }\textbf {\bibinfo {volume} {120}},\
  \bibinfo {pages} {103201} (\bibinfo {year} {2018})}\BibitemShut {NoStop}%
\bibitem [{\citenamefont {Hankin}\ \emph {et~al.}(2014)\citenamefont {Hankin},
  \citenamefont {Jau}, \citenamefont {Parazzoli}, \citenamefont {Chou},
  \citenamefont {Armstrong}, \citenamefont {Landahl},\ and\ \citenamefont
  {Biedermann}}]{PhysRevA.89.033416}%
  \BibitemOpen
  \bibfield  {author} {\bibinfo {author} {\bibfnamefont {A.~M.}\ \bibnamefont
  {Hankin}}, \bibinfo {author} {\bibfnamefont {Y.-Y.}\ \bibnamefont {Jau}},
  \bibinfo {author} {\bibfnamefont {L.~P.}\ \bibnamefont {Parazzoli}}, \bibinfo
  {author} {\bibfnamefont {C.~W.}\ \bibnamefont {Chou}}, \bibinfo {author}
  {\bibfnamefont {D.~J.}\ \bibnamefont {Armstrong}}, \bibinfo {author}
  {\bibfnamefont {A.~J.}\ \bibnamefont {Landahl}},\ and\ \bibinfo {author}
  {\bibfnamefont {G.~W.}\ \bibnamefont {Biedermann}},\ }\bibfield  {title}
  {\bibinfo {title} {Two-atom rydberg blockade using direct 6$s$ to $np$
  excitation},\ }\href {https://doi.org/10.1103/PhysRevA.89.033416} {\bibfield
  {journal} {\bibinfo  {journal} {Phys. Rev. A}\ }\textbf {\bibinfo {volume}
  {89}},\ \bibinfo {pages} {033416} (\bibinfo {year} {2014})}\BibitemShut
  {NoStop}%
\bibitem [{\citenamefont {Goy}\ \emph {et~al.}(1982)\citenamefont {Goy},
  \citenamefont {Raimond}, \citenamefont {Vitrant},\ and\ \citenamefont
  {Haroche}}]{PhysRevA.26.2733}%
  \BibitemOpen
  \bibfield  {author} {\bibinfo {author} {\bibfnamefont {P.}~\bibnamefont
  {Goy}}, \bibinfo {author} {\bibfnamefont {J.~M.}\ \bibnamefont {Raimond}},
  \bibinfo {author} {\bibfnamefont {G.}~\bibnamefont {Vitrant}},\ and\ \bibinfo
  {author} {\bibfnamefont {S.}~\bibnamefont {Haroche}},\ }\bibfield  {title}
  {\bibinfo {title} {Millimeter-wave spectroscopy in cesium rydberg states.
  quantum defects, fine- and hyperfine-structure measurements},\ }\href
  {https://doi.org/10.1103/PhysRevA.26.2733} {\bibfield  {journal} {\bibinfo
  {journal} {Phys. Rev. A}\ }\textbf {\bibinfo {volume} {26}},\ \bibinfo
  {pages} {2733} (\bibinfo {year} {1982})}\BibitemShut {NoStop}%
\bibitem [{\citenamefont {Beterov}\ \emph {et~al.}(2009)\citenamefont
  {Beterov}, \citenamefont {Ryabtsev}, \citenamefont {Tretyakov},\ and\
  \citenamefont {Entin}}]{PhysRevA.79.052504}%
  \BibitemOpen
  \bibfield  {author} {\bibinfo {author} {\bibfnamefont {I.~I.}\ \bibnamefont
  {Beterov}}, \bibinfo {author} {\bibfnamefont {I.~I.}\ \bibnamefont
  {Ryabtsev}}, \bibinfo {author} {\bibfnamefont {D.~B.}\ \bibnamefont
  {Tretyakov}},\ and\ \bibinfo {author} {\bibfnamefont {V.~M.}\ \bibnamefont
  {Entin}},\ }\bibfield  {title} {\bibinfo {title} {Quasiclassical calculations
  of blackbody-radiation-induced depopulation rates and effective lifetimes of
  rydberg $ns$, $np$, and $nd$ alkali-metal atoms with $n\ensuremath{\le}80$},\
  }\href {https://doi.org/10.1103/PhysRevA.79.052504} {\bibfield  {journal}
  {\bibinfo  {journal} {Phys. Rev. A}\ }\textbf {\bibinfo {volume} {79}},\
  \bibinfo {pages} {052504} (\bibinfo {year} {2009})}\BibitemShut {NoStop}%
\bibitem [{\citenamefont {Covey}\ \emph {et~al.}(2019)\citenamefont {Covey},
  \citenamefont {Madjarov}, \citenamefont {Cooper},\ and\ \citenamefont
  {Endres}}]{PhysRevLett.122.173201}%
  \BibitemOpen
  \bibfield  {author} {\bibinfo {author} {\bibfnamefont {J.~P.}\ \bibnamefont
  {Covey}}, \bibinfo {author} {\bibfnamefont {I.~S.}\ \bibnamefont {Madjarov}},
  \bibinfo {author} {\bibfnamefont {A.}~\bibnamefont {Cooper}},\ and\ \bibinfo
  {author} {\bibfnamefont {M.}~\bibnamefont {Endres}},\ }\bibfield  {title}
  {\bibinfo {title} {2000-times repeated imaging of strontium atoms in
  clock-magic tweezer arrays},\ }\href
  {https://doi.org/10.1103/PhysRevLett.122.173201} {\bibfield  {journal}
  {\bibinfo  {journal} {Phys. Rev. Lett.}\ }\textbf {\bibinfo {volume} {122}},\
  \bibinfo {pages} {173201} (\bibinfo {year} {2019})}\BibitemShut {NoStop}%
\bibitem [{\citenamefont {Beigman}\ and\ \citenamefont
  {Lebedev}(1995)}]{PhysRep.250.95}%
  \BibitemOpen
  \bibfield  {author} {\bibinfo {author} {\bibfnamefont {I.}~\bibnamefont
  {Beigman}}\ and\ \bibinfo {author} {\bibfnamefont {V.}~\bibnamefont
  {Lebedev}},\ }\bibfield  {title} {\bibinfo {title} {Collision theory of
  rydberg atoms with neutral and charged particles},\ }\href
  {https://doi.org/https://doi.org/10.1016/0370-1573(95)00074-Q} {\bibfield
  {journal} {\bibinfo  {journal} {Physics Reports}\ }\textbf {\bibinfo {volume}
  {250}},\ \bibinfo {pages} {95} (\bibinfo {year} {1995})}\BibitemShut
  {NoStop}%
\bibitem [{\citenamefont {Stebbings}\ and\ \citenamefont
  {Dunning}(1983)}]{RydStateAtMol}%
  \BibitemOpen
  \bibfield  {author} {\bibinfo {author} {\bibfnamefont {R.~F.}\ \bibnamefont
  {Stebbings}}\ and\ \bibinfo {author} {\bibfnamefont {F.~B.}\ \bibnamefont
  {Dunning}},\ }\href@noop {} {\emph {\bibinfo {title} {Rydberg States of Atoms
  and Molecules}}}\ (\bibinfo  {publisher} {Cambridge University Press},\
  \bibinfo {year} {1983})\BibitemShut {NoStop}%
\bibitem [{\citenamefont {Dunning}\ \emph {et~al.}(2009)\citenamefont
  {Dunning}, \citenamefont {Mestayer}, \citenamefont {Reinhold}, \citenamefont
  {Yoshida},\ and\ \citenamefont {Burgdörfer}}]{JPhysB.42.022001}%
  \BibitemOpen
  \bibfield  {author} {\bibinfo {author} {\bibfnamefont {F.~B.}\ \bibnamefont
  {Dunning}}, \bibinfo {author} {\bibfnamefont {J.~J.}\ \bibnamefont
  {Mestayer}}, \bibinfo {author} {\bibfnamefont {C.~O.}\ \bibnamefont
  {Reinhold}}, \bibinfo {author} {\bibfnamefont {S.}~\bibnamefont {Yoshida}},\
  and\ \bibinfo {author} {\bibfnamefont {J.}~\bibnamefont {Burgdörfer}},\
  }\bibfield  {title} {\bibinfo {title} {Engineering atomic rydberg states with
  pulsed electric fields},\ }\href
  {https://doi.org/10.1088/0953-4075/42/2/022001} {\bibfield  {journal}
  {\bibinfo  {journal} {Journal of Physics B: Atomic, Molecular and Optical
  Physics}\ }\textbf {\bibinfo {volume} {42}},\ \bibinfo {pages} {022001}
  (\bibinfo {year} {2009})}\BibitemShut {NoStop}%
\bibitem [{\citenamefont {Vriens}\ and\ \citenamefont
  {Smeets}(1980)}]{PhysRevA.22.940}%
  \BibitemOpen
  \bibfield  {author} {\bibinfo {author} {\bibfnamefont {L.}~\bibnamefont
  {Vriens}}\ and\ \bibinfo {author} {\bibfnamefont {A.~H.~M.}\ \bibnamefont
  {Smeets}},\ }\bibfield  {title} {\bibinfo {title} {Cross-section and rate
  formulas for electron-impact ionization, excitation, deexcitation, and total
  depopulation of excited atoms},\ }\href
  {https://doi.org/10.1103/PhysRevA.22.940} {\bibfield  {journal} {\bibinfo
  {journal} {Phys. Rev. A}\ }\textbf {\bibinfo {volume} {22}},\ \bibinfo
  {pages} {940} (\bibinfo {year} {1980})}\BibitemShut {NoStop}%
\bibitem [{\citenamefont {Flannery}\ and\ \citenamefont
  {McCann}(1979)}]{JPhysB.12.427}%
  \BibitemOpen
  \bibfield  {author} {\bibinfo {author} {\bibfnamefont {M.~R.}\ \bibnamefont
  {Flannery}}\ and\ \bibinfo {author} {\bibfnamefont {K.~J.}\ \bibnamefont
  {McCann}},\ }\bibfield  {title} {\bibinfo {title} {Systematic trends in the
  inelastic cross sections and form factors for nl→n'l' direct collisional
  transitions},\ }\href {https://doi.org/10.1088/0022-3700/12/3/020} {\bibfield
   {journal} {\bibinfo  {journal} {Journal of Physics B: Atomic and Molecular
  Physics}\ }\textbf {\bibinfo {volume} {12}},\ \bibinfo {pages} {427}
  (\bibinfo {year} {1979})}\BibitemShut {NoStop}%
\bibitem [{\citenamefont {MacAdam}\ \emph {et~al.}(1981)\citenamefont
  {MacAdam}, \citenamefont {Rolfes},\ and\ \citenamefont
  {Crosby}}]{PhysRevA.24.1286}%
  \BibitemOpen
  \bibfield  {author} {\bibinfo {author} {\bibfnamefont {K.~B.}\ \bibnamefont
  {MacAdam}}, \bibinfo {author} {\bibfnamefont {R.}~\bibnamefont {Rolfes}},\
  and\ \bibinfo {author} {\bibfnamefont {D.~A.}\ \bibnamefont {Crosby}},\
  }\bibfield  {title} {\bibinfo {title} {$l$ change in sodium rydberg atoms
  induced by ion collisions near the matching velocity},\ }\href
  {https://doi.org/10.1103/PhysRevA.24.1286} {\bibfield  {journal} {\bibinfo
  {journal} {Phys. Rev. A}\ }\textbf {\bibinfo {volume} {24}},\ \bibinfo
  {pages} {1286} (\bibinfo {year} {1981})}\BibitemShut {NoStop}%
\bibitem [{\citenamefont {Tielking}\ \emph {et~al.}(1995)\citenamefont
  {Tielking}, \citenamefont {Bensky},\ and\ \citenamefont
  {Jones}}]{PhysRevA.51.3370}%
  \BibitemOpen
  \bibfield  {author} {\bibinfo {author} {\bibfnamefont {N.~E.}\ \bibnamefont
  {Tielking}}, \bibinfo {author} {\bibfnamefont {T.~J.}\ \bibnamefont
  {Bensky}},\ and\ \bibinfo {author} {\bibfnamefont {R.~R.}\ \bibnamefont
  {Jones}},\ }\bibfield  {title} {\bibinfo {title} {Effects of imperfect
  unipolarity on the ionization of rydberg atoms by subpicosecond half-cycle
  pulses},\ }\href {https://doi.org/10.1103/PhysRevA.51.3370} {\bibfield
  {journal} {\bibinfo  {journal} {Phys. Rev. A}\ }\textbf {\bibinfo {volume}
  {51}},\ \bibinfo {pages} {3370} (\bibinfo {year} {1995})}\BibitemShut
  {NoStop}%
\bibitem [{\citenamefont {Gallagher}\ \emph {et~al.}(1977)\citenamefont
  {Gallagher}, \citenamefont {Humphrey}, \citenamefont {Cooke}, \citenamefont
  {Hill},\ and\ \citenamefont {Edelstein}}]{PhysRevA.16.1098}%
  \BibitemOpen
  \bibfield  {author} {\bibinfo {author} {\bibfnamefont {T.~F.}\ \bibnamefont
  {Gallagher}}, \bibinfo {author} {\bibfnamefont {L.~M.}\ \bibnamefont
  {Humphrey}}, \bibinfo {author} {\bibfnamefont {W.~E.}\ \bibnamefont {Cooke}},
  \bibinfo {author} {\bibfnamefont {R.~M.}\ \bibnamefont {Hill}},\ and\
  \bibinfo {author} {\bibfnamefont {S.~A.}\ \bibnamefont {Edelstein}},\
  }\bibfield  {title} {\bibinfo {title} {Field ionization of highly excited
  states of sodium},\ }\href {https://doi.org/10.1103/PhysRevA.16.1098}
  {\bibfield  {journal} {\bibinfo  {journal} {Phys. Rev. A}\ }\textbf {\bibinfo
  {volume} {16}},\ \bibinfo {pages} {1098} (\bibinfo {year}
  {1977})}\BibitemShut {NoStop}%
\bibitem [{\citenamefont {Paradis}\ \emph {et~al.}(2013)\citenamefont
  {Paradis}, \citenamefont {Zigo},\ and\ \citenamefont
  {Raithel}}]{PhysRevA.87.012505}%
  \BibitemOpen
  \bibfield  {author} {\bibinfo {author} {\bibfnamefont {E.}~\bibnamefont
  {Paradis}}, \bibinfo {author} {\bibfnamefont {S.}~\bibnamefont {Zigo}},\ and\
  \bibinfo {author} {\bibfnamefont {G.}~\bibnamefont {Raithel}},\ }\bibfield
  {title} {\bibinfo {title} {Highly polar states of rydberg atoms in strong
  magnetic and weak electric fields},\ }\href
  {https://doi.org/10.1103/PhysRevA.87.012505} {\bibfield  {journal} {\bibinfo
  {journal} {Phys. Rev. A}\ }\textbf {\bibinfo {volume} {87}},\ \bibinfo
  {pages} {012505} (\bibinfo {year} {2013})}\BibitemShut {NoStop}%
\bibitem [{\citenamefont {Pohl}\ \emph {et~al.}(2009)\citenamefont {Pohl},
  \citenamefont {Sadeghpour},\ and\ \citenamefont
  {Schmelcher}}]{PhysRep.484.181}%
  \BibitemOpen
  \bibfield  {author} {\bibinfo {author} {\bibfnamefont {T.}~\bibnamefont
  {Pohl}}, \bibinfo {author} {\bibfnamefont {H.}~\bibnamefont {Sadeghpour}},\
  and\ \bibinfo {author} {\bibfnamefont {P.}~\bibnamefont {Schmelcher}},\
  }\bibfield  {title} {\bibinfo {title} {Cold and ultracold rydberg atoms in
  strong magnetic fields},\ }\href
  {https://doi.org/https://doi.org/10.1016/j.physrep.2009.10.001} {\bibfield
  {journal} {\bibinfo  {journal} {Physics Reports}\ }\textbf {\bibinfo {volume}
  {484}},\ \bibinfo {pages} {181} (\bibinfo {year} {2009})}\BibitemShut
  {NoStop}%
\bibitem [{\citenamefont {Cidrim}\ \emph {et~al.}(2021)\citenamefont {Cidrim},
  \citenamefont {Pi\~neiro Orioli}, \citenamefont {Sanner}, \citenamefont
  {Hutson}, \citenamefont {Ye}, \citenamefont {Bachelard},\ and\ \citenamefont
  {Rey}}]{PhysRevLett.127.013401}%
  \BibitemOpen
  \bibfield  {author} {\bibinfo {author} {\bibfnamefont {A.}~\bibnamefont
  {Cidrim}}, \bibinfo {author} {\bibfnamefont {A.}~\bibnamefont {Pi\~neiro
  Orioli}}, \bibinfo {author} {\bibfnamefont {C.}~\bibnamefont {Sanner}},
  \bibinfo {author} {\bibfnamefont {R.~B.}\ \bibnamefont {Hutson}}, \bibinfo
  {author} {\bibfnamefont {J.}~\bibnamefont {Ye}}, \bibinfo {author}
  {\bibfnamefont {R.}~\bibnamefont {Bachelard}},\ and\ \bibinfo {author}
  {\bibfnamefont {A.~M.}\ \bibnamefont {Rey}},\ }\bibfield  {title} {\bibinfo
  {title} {Dipole-dipole frequency shifts in multilevel atoms},\ }\href
  {https://doi.org/10.1103/PhysRevLett.127.013401} {\bibfield  {journal}
  {\bibinfo  {journal} {Phys. Rev. Lett.}\ }\textbf {\bibinfo {volume} {127}},\
  \bibinfo {pages} {013401} (\bibinfo {year} {2021})}\BibitemShut {NoStop}%
\bibitem [{\citenamefont {Chew}\ \emph {et~al.}(2022)\citenamefont {Chew},
  \citenamefont {Tomita}, \citenamefont {Mahesh}, \citenamefont {Sugawa},
  \citenamefont {de~L{\'e}s{\'e}leuc},\ and\ \citenamefont
  {Ohmori}}]{NatPhoton.16.724}%
  \BibitemOpen
  \bibfield  {author} {\bibinfo {author} {\bibfnamefont {Y.}~\bibnamefont
  {Chew}}, \bibinfo {author} {\bibfnamefont {T.}~\bibnamefont {Tomita}},
  \bibinfo {author} {\bibfnamefont {T.~P.}\ \bibnamefont {Mahesh}}, \bibinfo
  {author} {\bibfnamefont {S.}~\bibnamefont {Sugawa}}, \bibinfo {author}
  {\bibfnamefont {S.}~\bibnamefont {de~L{\'e}s{\'e}leuc}},\ and\ \bibinfo
  {author} {\bibfnamefont {K.}~\bibnamefont {Ohmori}},\ }\bibfield  {title}
  {\bibinfo {title} {Ultrafast energy exchange between two single rydberg atoms
  on a nanosecond timescale},\ }\href
  {https://doi.org/10.1038/s41566-022-01047-2} {\bibfield  {journal} {\bibinfo
  {journal} {Nature Photonics}\ }\textbf {\bibinfo {volume} {16}},\ \bibinfo
  {pages} {724} (\bibinfo {year} {2022})}\BibitemShut {NoStop}%
\bibitem [{\citenamefont {Bharti}\ \emph {et~al.}(2023)\citenamefont {Bharti},
  \citenamefont {Sugawa}, \citenamefont {Mizoguchi}, \citenamefont {Kunimi},
  \citenamefont {Zhang}, \citenamefont {de~L\'es\'eleuc}, \citenamefont
  {Tomita}, \citenamefont {Franz}, \citenamefont {Weidem\"uller},\ and\
  \citenamefont {Ohmori}}]{2201.09590}%
  \BibitemOpen
  \bibfield  {author} {\bibinfo {author} {\bibfnamefont {V.}~\bibnamefont
  {Bharti}}, \bibinfo {author} {\bibfnamefont {S.}~\bibnamefont {Sugawa}},
  \bibinfo {author} {\bibfnamefont {M.}~\bibnamefont {Mizoguchi}}, \bibinfo
  {author} {\bibfnamefont {M.}~\bibnamefont {Kunimi}}, \bibinfo {author}
  {\bibfnamefont {Y.}~\bibnamefont {Zhang}}, \bibinfo {author} {\bibfnamefont
  {S.}~\bibnamefont {de~L\'es\'eleuc}}, \bibinfo {author} {\bibfnamefont
  {T.}~\bibnamefont {Tomita}}, \bibinfo {author} {\bibfnamefont
  {T.}~\bibnamefont {Franz}}, \bibinfo {author} {\bibfnamefont
  {M.}~\bibnamefont {Weidem\"uller}},\ and\ \bibinfo {author} {\bibfnamefont
  {K.}~\bibnamefont {Ohmori}},\ }\bibfield  {title} {\bibinfo {title}
  {Picosecond-scale ultrafast many-body dynamics in an ultracold
  rydberg-excited atomic mott insulator},\ }\href
  {https://doi.org/10.1103/PhysRevLett.131.123201} {\bibfield  {journal}
  {\bibinfo  {journal} {Phys. Rev. Lett.}\ }\textbf {\bibinfo {volume} {131}},\
  \bibinfo {pages} {123201} (\bibinfo {year} {2023})}\BibitemShut {NoStop}%
\bibitem [{\citenamefont {{Johnson}}(1972)}]{AstrophysJ.174.227}%
  \BibitemOpen
  \bibfield  {author} {\bibinfo {author} {\bibfnamefont {L.~C.}\ \bibnamefont
  {{Johnson}}},\ }\bibfield  {title} {\bibinfo {title} {{Approximations for
  Collisional and Radiative Transition Rates in Atomic Hydrogen}},\ }\href
  {https://doi.org/10.1086/151486} {\bibfield  {journal} {\bibinfo  {journal}
  {\apj}\ }\textbf {\bibinfo {volume} {174}},\ \bibinfo {pages} {227} (\bibinfo
  {year} {1972})}\BibitemShut {NoStop}%
\end{thebibliography}%

\end{document}